\documentstyle[12pt,epsf]{article}
\def\plotone#1{\centering \leavevmode
\epsfxsize=\columnwidth \epsfbox{#1}}

\def\refitem #1! #2! #3! #4! #5;{\hang\noindent
    \hangindent 36pt\makebox[36pt]{\hfil #1.~}\rm #2, 
    \it #3, \bf #4, \rm #5.\par}
\def\bookref #1! {\hang\noindent
    \hangindent 36pt\makebox[36pt]{\hfil #1.~}}

\def\etal   {{\sl et~al.}}
\def\wisk#1{\ifmmode{#1}\else{$#1$}\fi}
\def\percm  {\wisk{{\rm cm}^{-1}}}

\def\percmcub{\wisk{{\rm cm}^{-3}}}

\def\gsim   {\wisk{_>\atop^{\sim}}}

\def\um     {\wisk{{\rm \mu m}}}

\def\deg    {\wisk{^\circ\ }}

\def\COBE{{\sl COBE\/}}
\def\muK     {\wisk{{\rm \mu K}}}
\def\kmsMpc  {\wisk{{\rm km}\;{\rm s}^{-1}\;{\rm Mpc}^{-1}}}
\def\Amp     {\wisk{{\langle Q_{RMS}^2\rangle^{0.5}}}}

\catcode`\@=11
\long\def\@makefntext#1{ 
\protect\noindent \hbox to 3.2pt {\hskip-.9pt  
$^{{\ninerm\@thefnmark}}$\hfil}#1\hfill} 

 \def\@makefnmark{\hbox to 0pt{$^{\@thefnmark}$\hss}}  
	
\def\ps@myheadings{\let\@mkboth\@gobbletwo
\def\@oddhead{\hbox{} 
\rightmark\hfil\ninerm\thepage}   
\def\@oddfoot{}\def\@evenhead{\ninerm\thepage\hfil 
\leftmark\hbox{}}\def\@evenfoot{}
\def\sectionmark##1{}\def\subsectionmark##1{}}

\newcounter{sectionc}\newcounter{subsectionc}\newcounter{subsubsectionc}
\renewcommand{\section}[1] {\vspace{0.6cm}\addtocounter{sectionc}{1} 
\setcounter{subsectionc}{0}\setcounter{subsubsectionc}{0}\noindent 
	{\bf\thesectionc. #1}\par\vspace{0.4cm}}
\renewcommand{\subsection}[1] 
        {\vspace{0.6cm}\addtocounter{subsectionc}{1} 
	\setcounter{subsubsectionc}{0}\noindent 
	{\it\thesectionc.\thesubsectionc. #1}\par\vspace{0.4cm}}
\renewcommand{\subsubsection}[1] 
        {\vspace{0.6cm}\addtocounter{subsubsectionc}{1}
	\noindent {\rm\thesectionc.\thesubsectionc.\thesubsubsectionc. 
	#1}\par\vspace{0.4cm}}

\newcounter{appendixc}
\newcounter{subappendixc}[appendixc]
\newcounter{subsubappendixc}[subappendixc]

\renewcommand{\appendix}[1] {\vspace{0.6cm}
        \refstepcounter{appendixc}
        \setcounter{figure}{0}
        \setcounter{table}{0}
        \setcounter{equation}{0}
        \renewcommand{\thefigure}{\Alph{appendixc}.\arabic{figure}}
        \renewcommand{\thetable}{\Alph{appendixc}.\arabic{table}}
        \renewcommand{\theappendixc}{\Alph{appendixc}}
        \renewcommand{\theequation}{\Alph{appendixc}.\arabic{equation}}
        \noindent{\bf Appendix \theappendixc #1}\par\vspace{0.4cm}}

\newcounter{itemlistc}
\newcounter{romanlistc}
\newcounter{alphlistc}
\newcounter{arabiclistc}

\newcommand{\fcaption}[1]{
        \refstepcounter{figure}
        \setbox\@tempboxa = \hbox{\tenrm Fig.~\thefigure. #1}
        \ifdim \wd\@tempboxa > 6in
           {\begin{center}
        \parbox{6in}{\tenrm\baselineskip=12pt Fig.~\thefigure. #1 }
            \end{center}}
        \else
             {\begin{center}
             {\tenrm Fig.~\thefigure. #1}
              \end{center}}
        \fi}

\newcommand{\tcaption}[1]{
        \refstepcounter{table}
        \setbox\@tempboxa = \hbox{\tenrm Table~\thetable. #1}
        \ifdim \wd\@tempboxa > 6in
           {\begin{center}
        \parbox{6in}{\tenrm\baselineskip=12pt Table~\thetable. #1 }
            \end{center}}
        \else
             {\begin{center}
             {\tenrm Table~\thetable. #1}
              \end{center}}
        \fi}

\def\@citex[#1]#2{\if@filesw\immediate\write\@auxout
	{\string\citation{#2}}\fi
\def\@citea{}\@cite{\@for\@citeb:=#2\do
	{\@citea\def\@citea{,}\@ifundefined
	{b@\@citeb}{{\bf ?}\@warning
	{Citation `\@citeb' on page \thepage \space undefined}}
	{\csname b@\@citeb\endcsname}}}{#1}}

\newif\if@cghi
\def\cite{\@cghitrue\@ifnextchar [{\@tempswatrue
	\@citex}{\@tempswafalse\@citex[]}}
\def\citelow{\@cghifalse\@ifnextchar [{\@tempswatrue
	\@citex}{\@tempswafalse\@citex[]}}
\def\@cite#1#2{{$\null^{#1}$\if@tempswa\typeout
	{IJCGA warning: optional citation argument 
	ignored: `#2'} \fi}}

\def\fnt#1#2{\footnotetext{\kern-.3em
	{$^{\mbox{\sevenrm #1}}$}{#2}}}

 1
 1
 1

\font\tenrm=cmr10

\font\ninerm=cmr9

\begin{document}

\pagestyle{myheadings}

\markboth{}{}

\noindent
\makebox[0pt][l]{
\raisebox{36pt}[0pt][0pt]{Invited talk at the 23rd Intl. Cosmic
Ray Conf., Calgary, July 1993}}
\makebox[\hsize]{{\Large Cosmology/COBE}}

\vspace{0.2in}

\begin{center}
EDWARD L. WRIGHT\\
{\it UCLA Dept. of Astronomy}\\
{\it Los Angeles CA 90024-1562}
\end{center}

\begin{abstract}
\noindent
The results from the \COBE\footnote{
The National Aeronautics and Space Administration/Goddard Space Flight
Center (NASA/GSFC) is responsible for the design, development, and
operation of the Cosmic Background Explorer (COBE).
Scientific guidance is provided by the COBE Science Working Group.
GSFC is also responsible for the development of the analysis software and
for the production of the mission data sets.
}
satellite are in close quantitative
agreement with the predictions of the standard hot Big Bang model,
suggesting that the Universe was once hot, dense and isothermal,
giving a background radiation spectrum that is close to a
perfect blackbody.  The need to preserve this blackbody spectrum
to match the spectrum observed by the FIRAS instrument on \COBE\ places 
strong limits on events and
scenarios occurring later than 1 year after the Big Bang.
The observation of intrinsic anisotropy of the microwave background
by the DMR instrument on \COBE\ provides a measurement of the magnitude 
of the gravitational
potential fluctuations that existed $10^{5.5}$ years after the Big Bang.
These potential perturbations were either
produced in an inflationary epoch sometime during the first 
picosecond after the Big Bang, or else they are initial conditions
dating from $t = 0$.
The angular power spectrum of the correlations is compatible
with the inflationary scenario.
The observed amplitude of $\Delta T$ can be used to restrict theories of 
the fundamental
structure of matter on micro-physical scales and at energies higher than
that of the most energetic cosmic rays.
The gravitational forces implied by the observed potential can produce
the observed clustering of galaxies, but only if the density of the
Universe is dominated by some form of transparent ``dark'' matter,
which interacts weakly with photons.
While the FIRAS and DMR instruments have 7\deg beams, 
the third instrument on \COBE, DIRBE, has collected data with an
0.7\deg beam that will be
used to improve our knowledge of the cosmic infrared background, but
only after the strong foreground emissions from the interplanetary
dust cloud and galactic dust are modeled and removed.
\end{abstract}

\section{Introduction}

The observables of modern cosmology include the Hubble expansion of the 
Universe; the  ages of stars and clusters;  
the distribution and streaming motions of galaxies;
the content of the  Universe  (its mass density, composition, and the 
abundances of the light elements); 
the existence, spectrum and anisotropy of the cosmic microwave background 
radiation; and other potential backgrounds in the infrared, ultraviolet, 
x-ray, and gamma-ray spectral regions. 
The purpose of the \COBE\ mission is to make definitive
measurements of two of these observable cosmological fossils: the cosmic
microwave background (CMB) radiation and the cosmic infrared background
(CIB) radiation.  Since the  discovery  of the CMB$^1$,
many experiments have been performed to measure the CMB spectrum and 
spatial 
anisotropies over a wide range of wavelengths and angular scales.   Fewer
attempts have been made to conduct a sensitive search for a CIB radiation,
expected to  result  from the cumulative emissions of luminous objects
formed after the universe  cooled  sufficiently to permit the first stars
and galaxies to form.

The three scientific instruments on \COBE\ are the Far Infrared Absolute 
Spectrophotometer (FIRAS), the Differential Microwave Radiometers (DMR), 
and the Diffuse Infrared Background Experiment (DIRBE).  
The purpose of FIRAS 
is to make a precision measurement of the spectrum of the CMB from 1 cm 
to 100 $\mu$m.  The purpose of DMR is to search for variations in the
temperature of the CMB on
angular scales larger than 7$^\circ$ at frequencies of 31.5, 53, 
and 90 GHz.   
The purpose of DIRBE is to search for a CIB by making absolute brightness 
measurements of the diffuse infrared radiation in 10 photometric bands 
from 1 
to 240 $\mu$m and polarimetric measurements from 1 to 3.5 $\mu$m. 
The FIRAS and DIRBE instruments are located inside a 650 liter superfluid 
liquid helium dewar.
See Boggess \etal$^2$ for a full description of the {\it COBE} mission.


The \COBE\ mission is the product of many years of work by a large team of
scientists and engineers.  Credit for all of the results presented in this
paper must be shared with the other members of the the \COBE\ Science
Working Group: Chuck Bennett, Ed Cheng, Eli Dwek, Mike Hauser, Tom Kelsall,
John Mather, Harvey Moseley, Nancy Boggess, Rick Shafer, Bob Silverberg, 
George Smoot, Steve Meyer, Rai Weiss, Sam Gulkis, Mike Janssen, 
Dave Wilkinson, Phil Lubin, and Tom Murdock.  
Bennett \etal$^3$ is a review of the history of the \COBE\
project and its results up to the discovery of the anisotropy by the DMR,
but not including the latest FIRAS limits on distortions.

Some of the members of this team have been working on \COBE\ since the
release of the NASA Announcement of Oppurtunities AO-6 and AO-7 in 1974, 
when the proposals for what became the \COBE\ project were submitted.  
I have been working on \COBE\ since the beginning of 1978.
\COBE\ was successfully launched at 14:34 UT on 18 November 1989 
from Vandenberg AFB, California, and
returned high quality scientific data from all three instruments for 
ten months until the liquid helium ran out.  But about 50\% of the 
instruments do not require liquid helium, and are still returning valuable
scientific data in July 1993.
While the \COBE\ mission has been successful,
making the ``discovery of the century'', 
one must remember that this work is
based on the earlier work (in the $20^{th}$ century!) of Hubble$^4$
and Penzias and Wilson$^1$, who discovered the expansion of the Universe 
and the microwave background itself.  As a consequence of these two 
discoveries, one knows that the early Universe was hot and dense.  
When the density and
temperature are high, the photon creation and destruction rates are high,
and are sufficient to guarantee the formation of a good blackbody
spectrum.  Later, as the Universe expands and cools, 
the photon creation and
destruction rates become much slower than the expansion 
rate of the Universe,
which allow distortions of the spectrum to survive.  
In the standard model the
time from which distortions could survive is 1 year after the Big Bang at
a redshift $z \approx 10^{6.4}$.

However, the expansion of the Universe will not produce a distortion by
itself, since adiabatically expanding a blackbody 
radiation field results in another blackbody with a lower temperature.  
Thus the existence of a distorted spectrum in the hot Big Bang
model requires the existence at time later than 1 year after the Big Bang
of both an energy source and an emission
mechanism that can produce photons that are now in the millimeter spectral 
range.  Conversely, a lack of distortions can be used to place limits
on any such energy source, such as decaying neutrinos, dissipation of
turbulence, etc.

At a time $10^{5.4}$ years\footnote{
With $H_\circ = 50,\;\Omega = 1$.}
after the Big Bang, at $z \approx 1360$, the
temperature has fallen to the point where helium and then hydrogen 
have 50\% (re)combined into transparent gases$^5$.
The surface of last scattering ($\partial \tau/\partial \ln(1+z) = 1$)
occurs later, at $z \approx 1160$ or $10^{5.7}$ years after the
Big Bang.
The electron scattering which had impeded the free
motion of the CMB photons until this epoch is removed, 
and the photons stream
across the Universe.  Before recombination, 
the radiation field at any point
was constrained to be nearly isotropic because the rapid scattering
scrambled the directions of photons.  
The radiation field was not required to be homogeneous, 
because the photons remained approximately fixed in comoving
co-ordinates.  
After recombination, the free streaming of the photons has the
effect of averaging the intensity of the microwave background 
over a region
with a size equal to the horizon size.  Thus after recombination any
inhomogeneity in the microwave background spectrum is smoothed out.  
Note that this
inhomogeneity is not lost: instead, it is converted into anisotropy.  
When we
study the isotropy of the microwave background, we are looking back to the
surface of last scattering $\approx 10^{5.5}$ years after the Big Bang.  
But the hot spots
and cold spots we are studying existed as inhomogeneities in the Universe
before recombination.  
Since the $7^\circ$ beam used by the DMR instrument on
\COBE\ is larger than the horizon size at recombination, 
these inhomogeneities cannot be constructed in a causal fashion during 
the epoch before recombination in the standard Big Bang model.   
Instead, they must be installed ``just so'' in the initial conditions.  
In the inflationary scenario$^6$ these large scale structures were once 
smaller than the horizon size during the inflationary epoch, 
but grew to be much larger than the horizon.  Causal
physics acting $10^{-35}$~seconds after the Big Bang can produce the
large-scale inhomogeneities studied by the DMR.

\section{COBE Orbit and Attitude}

\COBE\ was launched at dawn on 18 November 1989 into a sun-synchronous
orbit with an inclination of $99^\circ$ and an altitude of 900 km.
By choosing a suitable combination of inclination and altitude, the 
precession rate of the orbit can be set to follow the motion of the Sun
around the sky at 1 cycle per year.  Thus the line perpendicular to the
orbit plane always points approximately toward the Sun.
Since \COBE\ was launched toward the South in the morning, 
the end of the orbit normal which points toward the Sun is at declination
$-9^\circ$.  At the June solstice, when the declination of the Sun is 
$+23^\circ$, there is a $32^\circ$ deviation from the ideal situation
where the Earth-Sun line is exactly perpendicular to the orbit.  Since
the depression of the horizon given by $\cos d = R_\oplus/a$,
where $a$ is the radius of the orbit of \COBE, is $29^\circ$,
there are eclipses when \COBE\ passes over the South pole for a period of
two months centered on the June solstice.  During the same season at the
North pole the angle between the Sun and anti-Sunward limb of the Earth
is only $177^\circ$.  It is thus not possible to keep both Sunlight and
Earthshine out of the shaded cavity around the dewar.  Since the Sun
produces a much greater bolometric intensity than a thin crescent of
Earth limb, the choice to keep the Sun below the plane of the shade is
an obvious one.

The pointing of \COBE\ is normally set to maintain the angle between
the spin axis and the Sun at $94^\circ$.  This leaves the Sun
$4^\circ$ below the plane of the shade.  During eclipse seasons
this margin is shaved to $2^\circ$ to reduce the amount of Earthshine
into the shaded cavity where the instruments sit.  The angle between
the plane of the shade and the Sun is known as the ``roll'' angle.
The ``pitch'' angle describes the rotation of the spin axis around
the Earth-Sun line.  This rotation is programmed for a constant
angular rate of 1 cycle per orbit.  The nadir is set to be close
to the minus spin axis, but because the Sun can be up to $32^\circ$
from the orbit normal and the Sun is kept $4^\circ$ below the plane of
the shade, the spin axis can be up to $36^\circ$ away
from the zenith.  The uniform rate in pitch allows the nadir
to oscillate by up to $\pm 6^\circ$ relative to the spin-Sun plane
with a period of 2 cycles per orbit.  This motion is similar to the
effect of the inclination of the ecliptic on the equation of time
that can be seen in an analemma.  To keep the ``wind'' caused by the
orbital motion of \COBE\ from impinging on the cryogenic optics, the
pitch is biased backward by $6^\circ$.  The ``yaw'' motion of \COBE\ 
is its spin.  The spin period is about 73 seconds.

There are two types of instrument pointings on \COBE.  
The DMR and DIRBE instruments
have fields-of-view located $30^\circ$ away from the spin axis.  As a
result they scan cycloidal paths through the sky, that cover the band
from $64^\circ$ to $124^\circ$ from the Sun.  This band contains 
50\% of the sky, and it is completely covered in a day by the DMR or
DIRBE.
This rapid coverage of the sky implies a rapid scan rate:
$2.5^\circ/{\rm sec}$.  
The data rates are also high: the DIRBE transmits
8 readings/channel/sec, while the DMR transmits 2 readings/channel/sec.
The FIRAS
points straight out along the spin axis.  As a result its field-of-view
moves only at the orbital rate of $3.6^\circ/{\rm min}$.  
Since FIRAS coadds interferograms over a collection period of about 
30 to 60 seconds, this slow scanning of the sky is essential.  

Over the course of a year the Sun moves around the sky, and the scan
paths of the instruments follow this motion.  The leading edge of the
wide scanned band for the DIRBE and DMR caught up to the trailing
edge of the band at start of mission after 4 months, leading to
100\% sky coverage.  
For the FIRAS total sky coverage would in
principle be obtained after 6 months, but because of gaps in the
observations caused by calibration runs and lunar interference,
the actual sky coverage for FIRAS, defined as having the center of
the beam within a pixel, is about 90\%.  The coverage gaps
are sufficiently narrow, however, that every direction on the
sky was covered by at least the half power contour of the beam.

The attitude determination system for \COBE\ uses infrared star
sightings from DIRBE to correct a gyro propagated attitude solution.
This method requires 3 working gyros that are not co-planar.
On 10 July 1993, \COBE\ suffered its fourth gyro failure
out of six gyros,
which will degrade the accuracy of the attitude solutions
by about a factor of 5, to a $1\sigma$ accuracy of about $0.1^\circ$.
(One of the failed gyros is noisy but operable.)
This accuracy will be sufficient for the DMR experiment with its
7\deg beam, provided that the errors are not systematic.
Current planning is to operate \COBE\ until December 1993, which
will complete four years of observations with the DMR and the
short wavelength channels of the DIRBE.

\section{DIRBE Measurements}

The Diffuse Infrared Background experiment (DIRBE) on \COBE\ is
designed to search for a cosmic infrared background (CIB).  
Models of the CIB
radiation take light from distant, unresolved galaxies, and redshift
it into the near infrared bands$^7$;
or absorb the starlight on dust,
reradiate it in the far-infrared, and then redshift it into the
sub-millimeter band$^8$.
These models predict fluxes that are approximately two orders of magnitude
lower than the emission from the Milky Way, which makes detection of the
CIB a difficult task, unless observations are made at wavelengths
where the redshift carries the energy from distant galaxies into a region
with little Milky Way emission.
The interplanetary dust cloud, which scatters sunlight to give the zodiacal
light, absorbs much more power than it scatters and reradiates this power
in the mid-infrared.
Thus from 5 \um\ to 100 \um\ the CIB is overwhelmed when observing from
a distance 1 au from the Sun.
However, there are windows between the scattered and thermal zodiacal light
at 3.5 \um\ and between the interstellar dust in the Milky Way and the
CMBR at 200-300 \um\ where the CIB may be detectable.

The DIRBE instrument is an off-axis Gregorian telescope with a 19 cm 
diameter
primary.  The Gregorian design has an intermediate image of the sky, which
allows for the placement of a stray-light stop between the primary and the
secondary, and the secondary forms an image of the primary where a second
stray-light stop is located.  Baffle vanes in the telescope tube absorb
radiation coming from far off-axis, so the main path for stray light 
involves
scattering on the primary mirror when it is illuminated by a slight 
off-axis
source.  Great care was taken to avoid contamination and to preserve the
cleanliness of the optics, so the off-axis response of the DIRBE is very 
small.
The DIRBE beam is an $0.7^\circ \times 0.7^\circ$ square.  
Dichroics and aperture
division are used to gives simultaneous measurements in 10 spectral bands
in the same field of view.
The 10 bands in DIRBE are the ground-based photometry bands
J[1.2], K[2.3], L[3.4], and M[4.9];
the four IRAS bands at 12, 25, 60, and 100 $\mu$m;
and two long wavelength bands at 140 and 240 $\mu$m.
The shortest wavelengths (JKL) bands, 
which observe scattered light from the
interplanetary dust cloud, have 3 detectors per band with polarizers to
measure linear polarization.
A tuning fork running at 32 Hz serves as a chopper for the DIRBE.  This
gives a well-determined zero level by comparing the sky signal to the
2~K interior of the DIRBE, 
allowing the search for an isotropic CIB
to proceed.  A shutter blocks the optical path less frequently, so the
accuracy of the zero level can be checked.  
When the shutter is in the beam,
its back side is a mirror which reflect light from a commandable 
internal reference source (IRS) into the detectors, providing a gain
calibration several times per orbit.
During the 10 months of cryogenic operation, the achieved
instrument rms sensitivity per  field of view was $\lambda
I(\lambda) = (1.0,\;0.9,\;0.6,\;0.5,\;0.3,\;0.4,\;0.4,\;0.1,\;11.0,$ $4.0)
\times 10^{-9}$ W m$^{-2}$ sr$^{-1}$, respectively for the ten wavelength 
bands listed above.  These noise levels are considerably smaller
than the total fluxes measured by DIRBE at the South Ecliptic Pole given
in Table \ref{sep}.
Note that the predicted extragalactic flux
is $3 \times 10^{-8}$ W\,m$^{-2}$sr$^{-1}$
at both 3.5 \um$^7$ and 240 \um$^8$, which is 20\% of the total 
SEP flux at 3.5 \um\ and 43\% at 240 \um.

\begin{table}[tb]
\begin{center}
\begin{tabular}{|l|r@{.}l|r@{.}l@{~}c@{~}l|r|}
\hline
Reference & \multicolumn{2}{c|}{$\lambda$}      &
\multicolumn{4}{c|}{$\lambda I_\lambda$} & $T_B(\lambda)$ \\
          & \multicolumn{2}{c|}{($\mu{\rm m}$)} &
\multicolumn{4}{c|}{($10^{-7}$ W m$^{-2}$ sr$^{-1}$)} & K \\
\hline
DIRBE                        &    1&2   &    8&3&$\pm$&~3.3 &  351 \\
{\it (South Ecliptic Pole)}  &    2&3   &    3&5&$\pm$&~1.4 &  216 \\
                             &    3&4   &    1&5&$\pm$&~0.6 &  140 \\
                             &    4&9   &    3&7&$\pm$&~1.5 &  109 \\
                             &   12&     &    29&&$\pm$&12  &  56 \\
                             &   22&     &    21&&$\pm$&~8  &  34  \\
                             &   55&     &     2&3&$\pm$&~1 &  15  \\
                             &   96&     &     1&2&$\pm$&~0.5 & 9.2 \\
                             &  151&     &     1&3&$\pm$&~0.7 & 6.6 \\
                             &  241&     &     0&7&$\pm$&~0.4 & 4.5 \\
\hline
\end{tabular}
\end{center}
\caption{Brightness and Planck Brightness Temperature of the Diffuse
Infrared Sky}
\label{sep}
\end{table}

The final determination of the magnitude of the extragalactic background
will depend on modeling and removing the zodiacal light.  The DIRBE
beam is aligned 30\deg from the spin axis of \COBE, so it swings
from 64\deg elongation to 124\deg elongation in every rotation of the
satellite.   In principle, the variation of intensity with elongation
can be used to determine the magnitude of the zodiacal emission,
just as the variation of atmospheric extinction with secant of the
zenith angle allows one to determine the flux hitting the top of the
atmosphere.  However, the shape of the atmosphere is well known, while the
shape of the zodiacal dust cloud is poorly determined.
If the cloud is in equilibrium, then Jeans' theorem states that the
phase space density should be a function of the integrals of the motion:
total angular momentum, $L$, 
the z component of the angular momentum, $L_z$, and the energy, $E$.  
The total angular momentum and energy define the radius 
and eccentricity of an orbit.  If the eccentricity distribution can be 
ignored, one is left with a dependence on radius.  
The ratio of $L_z$ to $L$
defines the inclination of the orbit.  Thus the cloud should be
describable by a function $f(r,i)$, which suggests that {\it fan-shaped}
models should be used.  While $f(r,i)$ allows for the ``bands'' seen by
IRAS, the comet trails are clearly non-equilibrium structures outside
the scope of these models.
Moreover, the center from which $r$ is measured
and the axis from which $i$ is measured depend on 
planetary perturbations$^9$ and are both functions of $r$.
Clearly a lot of effort will be needed to achieve the hoped for
removal of the zodiacal light to an accuracy of 1\%
of the peak emission.

After the depletion of the liquid helium, the DIRBE JKLM channels continued
to operate.  The interior temperature of the Dewar 
appears to have stabilized
at $\approx 55$~K, which is adequate for the InSb detectors used in these
bands.  
While the ``blanked-off'' noise in these channels has risen due to the
increased temperature, the noise while observing the sky is dominated
by the confusion noise due to the multitude of faint stars.
These ``warm'' data have been used to provide star sightings for the
attitude determination.

\section{DMR Observations}

The Differential Microwave Radiometers (DMR) experiment on \COBE\ is
designed to measure small temperature differences 
from place to place on the sky.  
The DMR consists of three separate units, 
one for each of the three frequencies of
31.5, 53 and 90 GHz.  The field of view of each unit consists of two beams
that are separated by a $60^\circ$ angle that is bisected by the spin axis.
Each beam has a $7^\circ$ FWHM.  The DMR is only sensitive to the
brightness difference between these two beams.  This differencing is
performed by a ferrite waveguide switch that connects the receiver input
to one horn and then the other at a rate of 100 Hz.  The 
signal then goes through a mixer, an IF amplifier and a video detector.
The output of the video detector is demodulated by a lock-in amplifier
synchronized to the input switch.  The difference signal that results
is telemetered to the ground every 0.5 seconds.
Each radiometer has two channels: A and B.  
In the case of the 31.5 GHz radiometer, 
the two channels use a single pair of horns in opposite senses of circular
polarization.  In the 53 and 90 GHz radiometers there are 4 horns, and all
observe the same sense of linear polarization.

\begin{figure}[tb]
\plotone{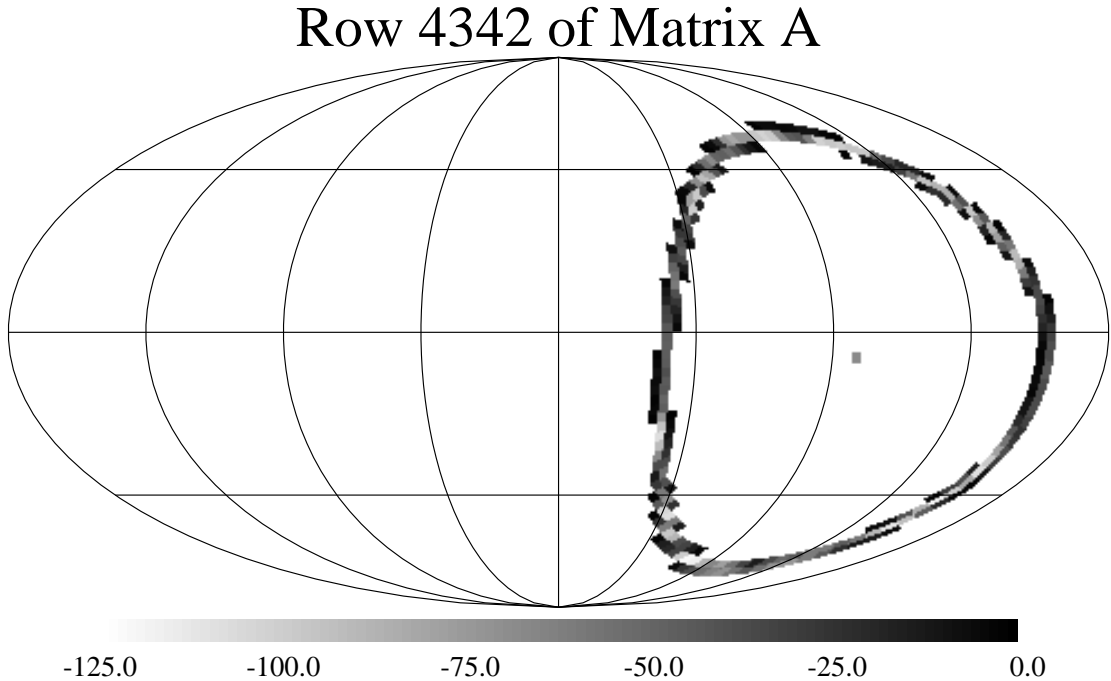}
\caption{The column through pixel 4342 of the matrix A for the 53 GHz
A channel.  The central pixel contains the galactic center, and is
saturated with a value of 10,599.  Ecliptic coordinates are used for
this plot.}
\label{A4342}
\end{figure}

\begin{figure}[tb]
\plotone{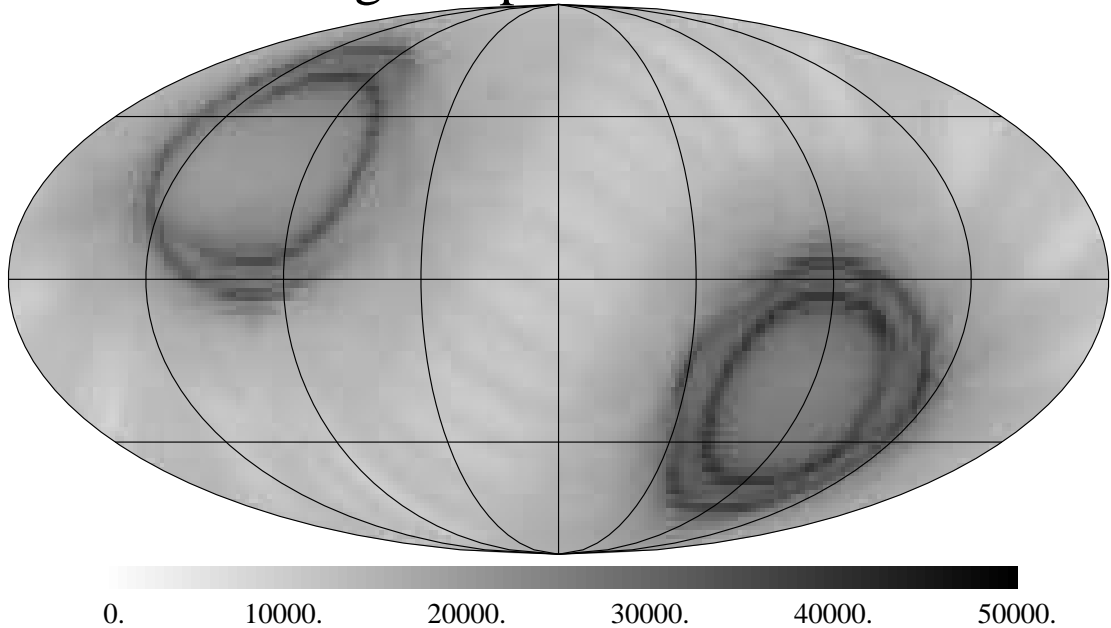}
\caption{The diagonal of the matrix A for the 53 A channel, otherwise known
as the coverage map, in galactic coordinates.}
\label{DIAG}
\end{figure}

The basic problem in the DMR data analysis is to construct a map of the
sky using the $6 \times 10^7$ differences that are collected each year
in each of the six channels.
There are only about $10^3$ beam areas on the sky, so the problem is
highly over-determined.  We have chosen to analyze the data using
6,144 pixels to cover the sky.  These are approximately equal area, and
arranged in a square grid of $32\times 32$ pixels on each of the 6 faces 
of a cube.  Within each pixel we assume that the temperature is constant,
so we are modeling the sky with a staircase function.  Therefore the
basic problem can be represented as a least-squares problem with 
$6\times 10^7$ equations in 6,144 variables.
One way to solve this problem is to average all of the observations
of each pixel.
This gives the temperature of the pixel relative to its
``reference ring'' of pixels 60\deg away.
Figure \ref{A4342} shows the reference ring for pixel 4342 which
contains the galactic center.
Then correct the average for each pixel by the weighted mean of the
temperatures in the reference ring.
The correction must be iterated using the corrected temperatures
until the process converges, but the convergence is slow.
In practice, the DMR maps are computed by inverting the
$6144 \times 6144$ matrix of normal equations.
Fortunately this matrix $A_{ij}$ is sparse and symmetric.
Thus only $9 \times 10^5$ elements need to be kept.
Figure \ref{A4342} can be viewed as a map made by plotting
$A_{ij}$ at the position of the $i^{th}$ pixel while $j$ is fixed
at $j = 4342$.  The diagonal $A_{ii}$ shown in Figure \ref{DIAG}
shows the number of observations per pixel.
This figure is plotted in galactic coordinates
to better show the structure of the heavily covered rings 26\deg
and 34\deg from the ecliptic poles.  The coverage near the North ecliptic
pole is reduced because the Earth limb rises above the sun shade over
the North pole during the May-July eclipse season.

\section{DMR Results}

\begin{figure}[tb]
\plotone{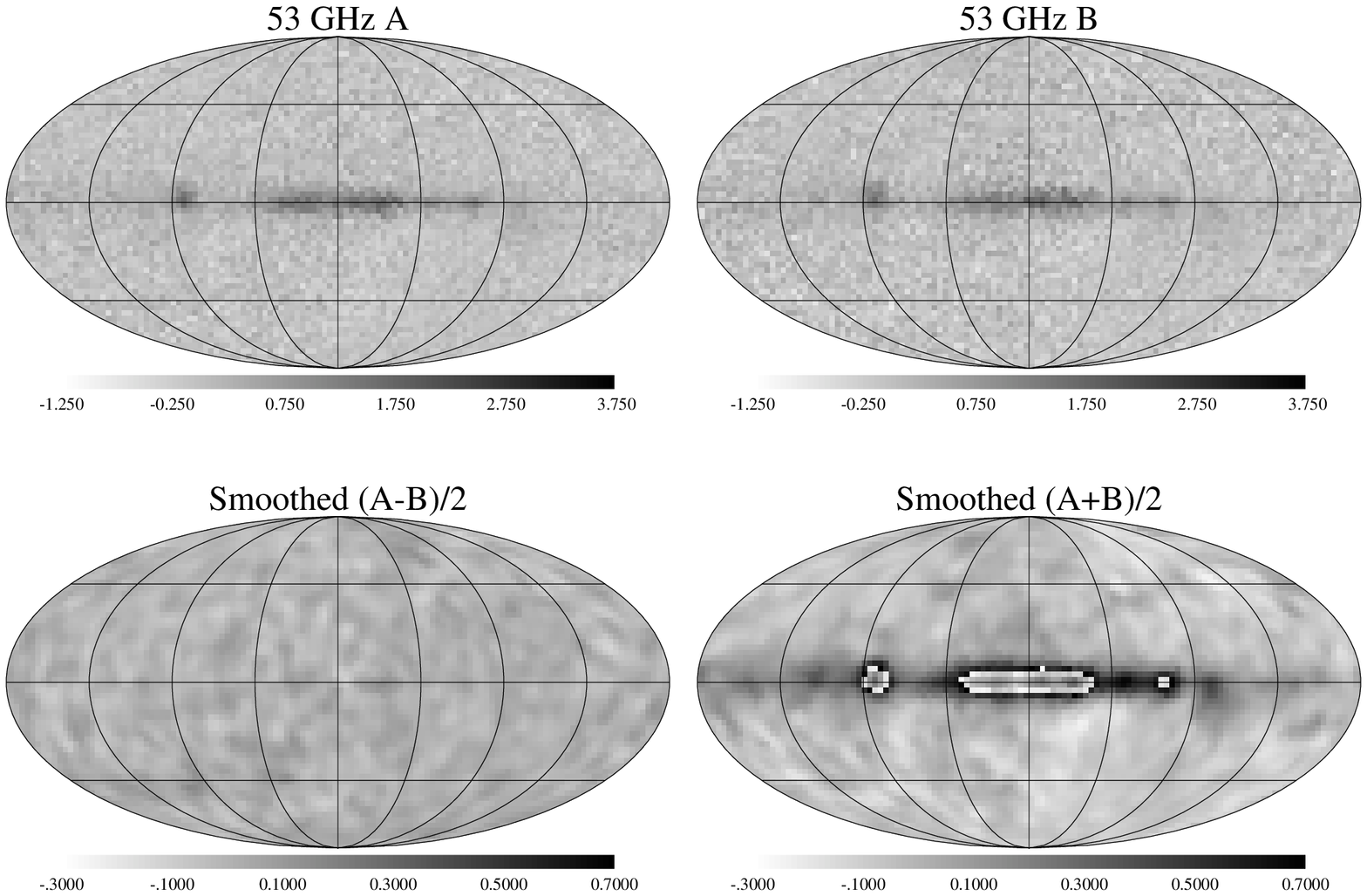}
\caption{The unsmoothed 53 GHz A \& B maps from the initial COBE data 
products, and the smoothed sum and difference maps.}
\label{maps53y1}
\end{figure}

The basic DMR result is the discovery
of an intrinsic anisotropy$^{10}$  of the microwave background, beyond the
dipole anisotropy$^{11,12}$.  This anisotropy,
when a monopole and dipole fit to $|b| > 20^\circ$ is removed from
the map, and the map is then smoothed to a resolution of 
$\approx 10^\circ$, is 30 \muK.  The correlation function of this
anisotropy is well fit by the expected correlation function for
the Harrison-Zeldovich$^{13,14,15}$
spectrum of primordial density perturbations
predicted by the inflationary scenario, and the amplitude is
consistent with many models of structure formation$^{16}$.
Galactic emission$^{17}$, systematic errors$^{18}$ and known
extra-galactic sources$^{19}$ cannot explain the observed signal.

\begin{figure}[tb]
\plotone{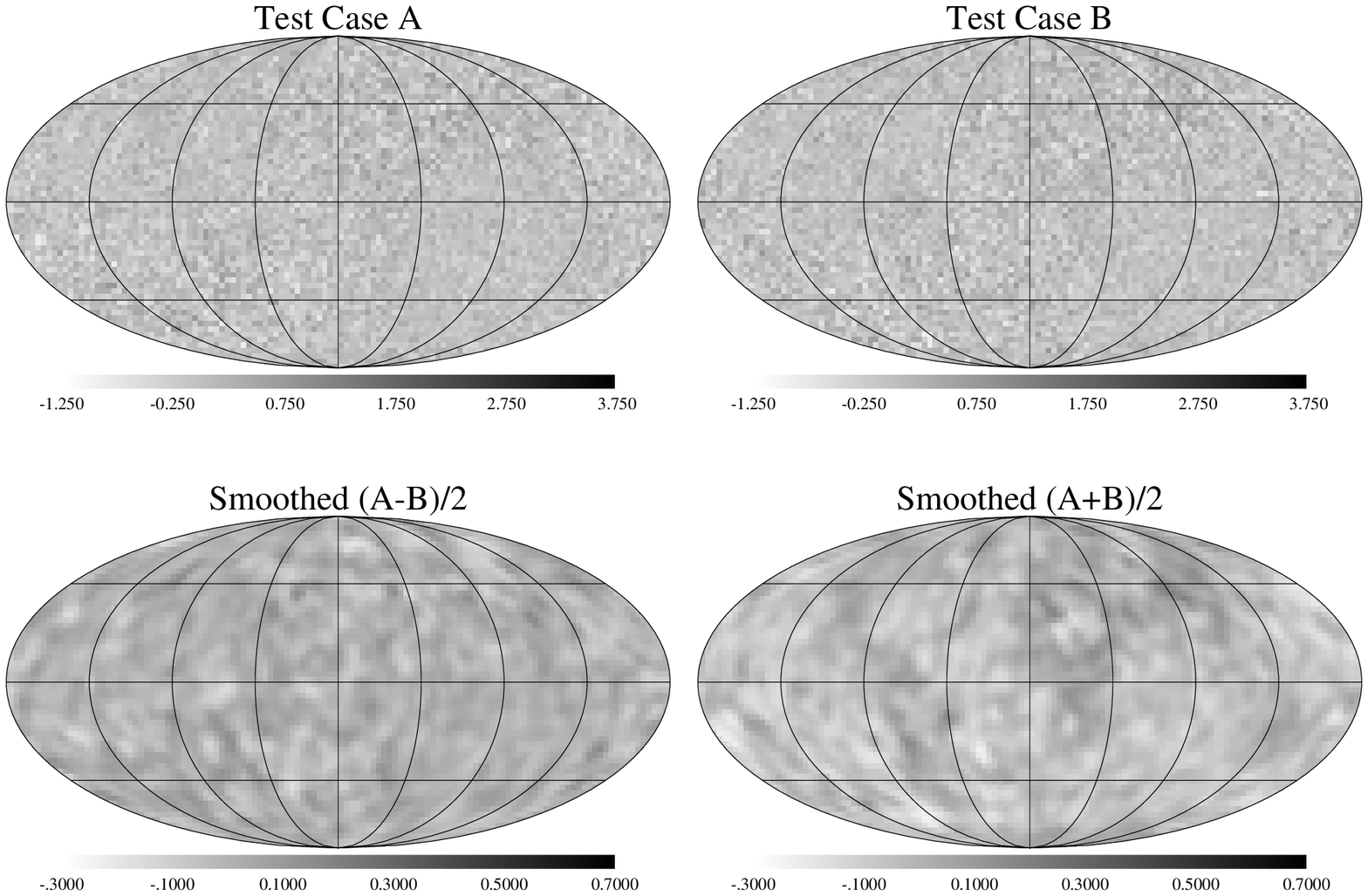}
\caption{The unsmoothed test case A \& B maps made by degrading a known
pattern to the DMR resolution and signal to noise ratio,
and the smoothed sum and difference maps.}
\label{testmaps}
\end{figure}

\begin{figure}[tb]
\plotone{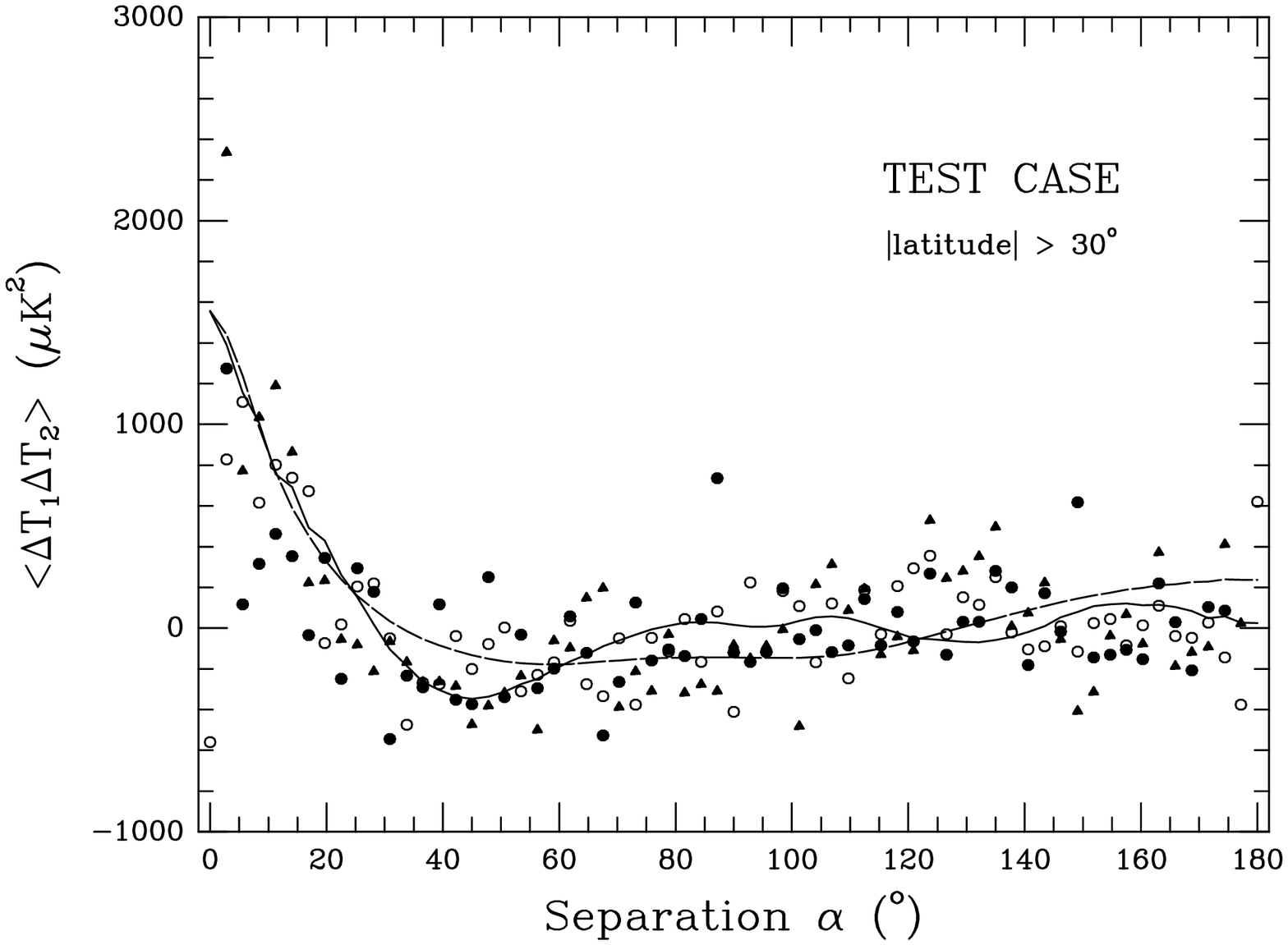}
\caption{The auto-correlations of the test case A \& B maps (solid dots),
the cross-correlation (open circles), and the correlation function of
the known input pattern at DMR resolution.  Also shown as a dashed curve
is the correlation function of an $n = 1$ Harrison-Zeldovich model.}
\label{testcc}
\end{figure}

\begin{figure}[tb]
\plotone{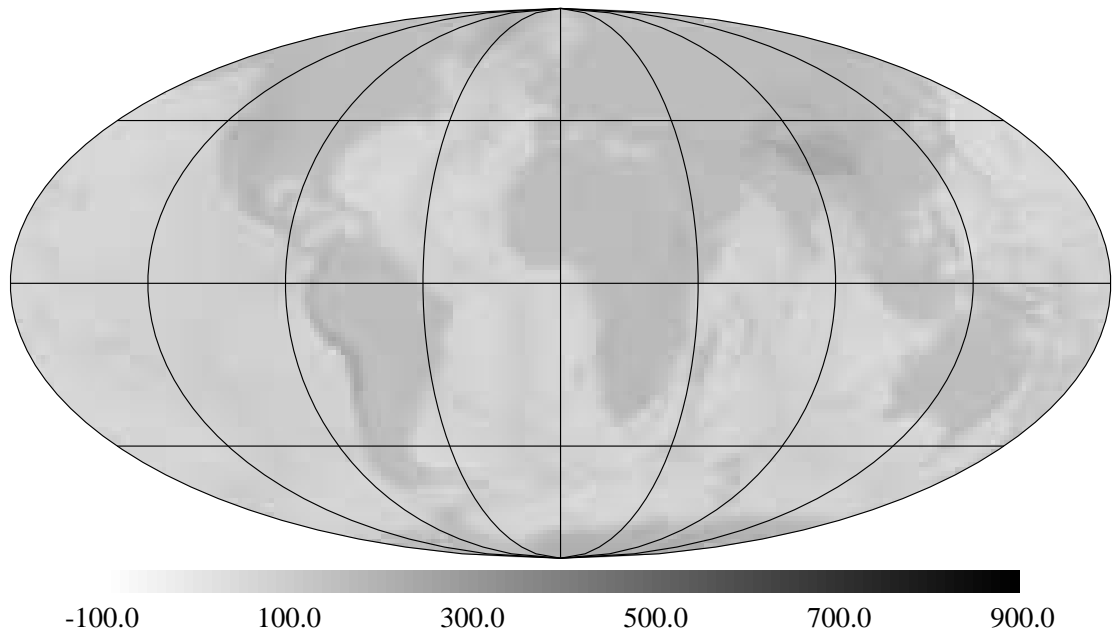}
\caption{The known pattern used to generate the test case maps.}
\label{earth}
\end{figure}

Even though the noise per pixel is much larger than the cosmic anisotropy,
the detection of anisotropy has a high statistical confidence level 
because of the large number of pixels in the DMR maps.  To illustrate this 
point, Figure \ref{maps53y1} shows the unsmoothed A and B side 53 GHz maps 
from the initial data products, plus the smoothed difference and sum maps.
Compare these to a test case with a known map, degraded to a comparable
signal to noise ratio, shown in Figure \ref{testmaps}.  Clearly a pattern
of some sort is starting to emerge in the smoothed sum map.
Figure \ref{testcc} shows the two auto-correlations and 
the cross-correlation
that can be computed from the unsmoothed test case maps.  The solid curve
in this figure is the known correlation function based on the known input 
map.
(A noiseless high resolution picture of the input map is in 
Figure \ref{earth}.)
Clearly the DMR maps show statistically significant correlated structure.
The dashed curve is the correlation function of a Harrison-Zeldovich
model with equal power on all scales, smoothed to the DMR resolution, 
with the amplitude adjusted to match the test case.  
The relatively good agreement between the dashed curve and the test case
correlation function shows that the good agreement between the DMR
correlation function and the Harrison-Zeldovich model$^{10}$,
while supporting inflation, does not prove inflation.

\begin{figure}[tb]
\plotone{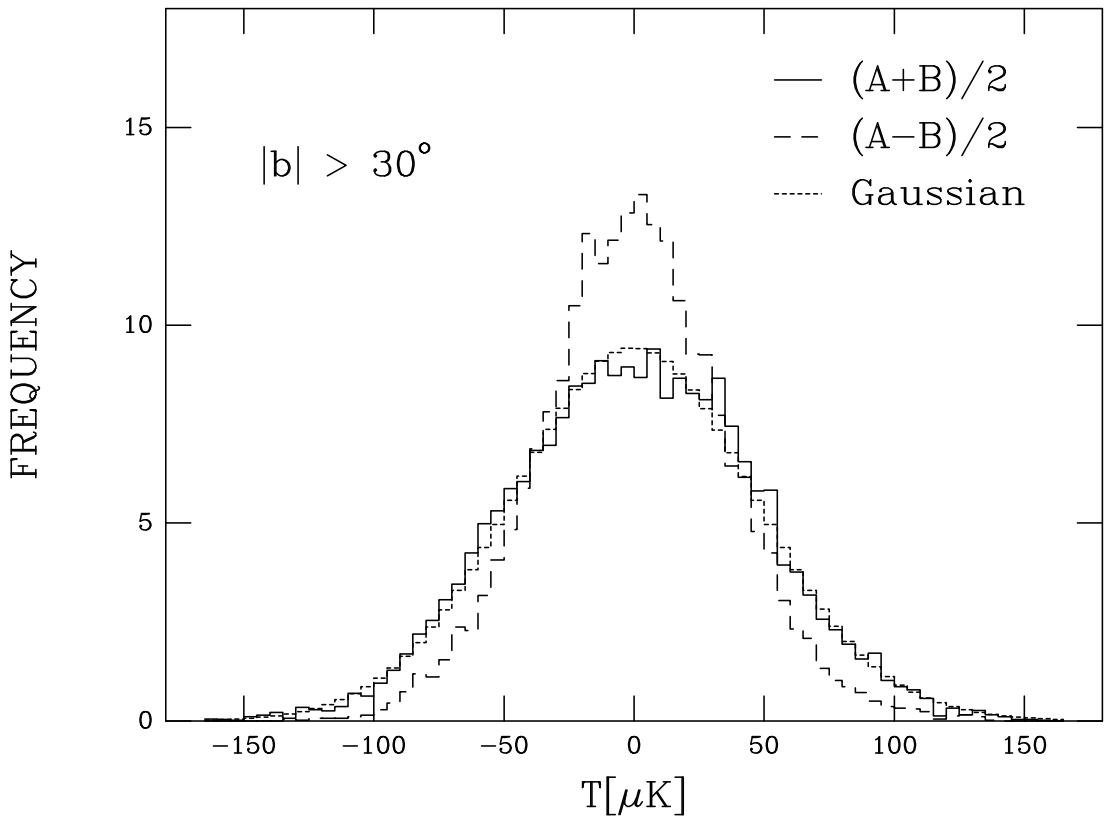}
\caption{The histogram of the smoothed 1 year 53+90 GHz (A-B)/2 map,
its convolution by a Gaussian, which matches the observed histogram
of the (A+B)/2 map.}
\label{hist}
\end{figure}

\begin{figure}[tb]
\plotone{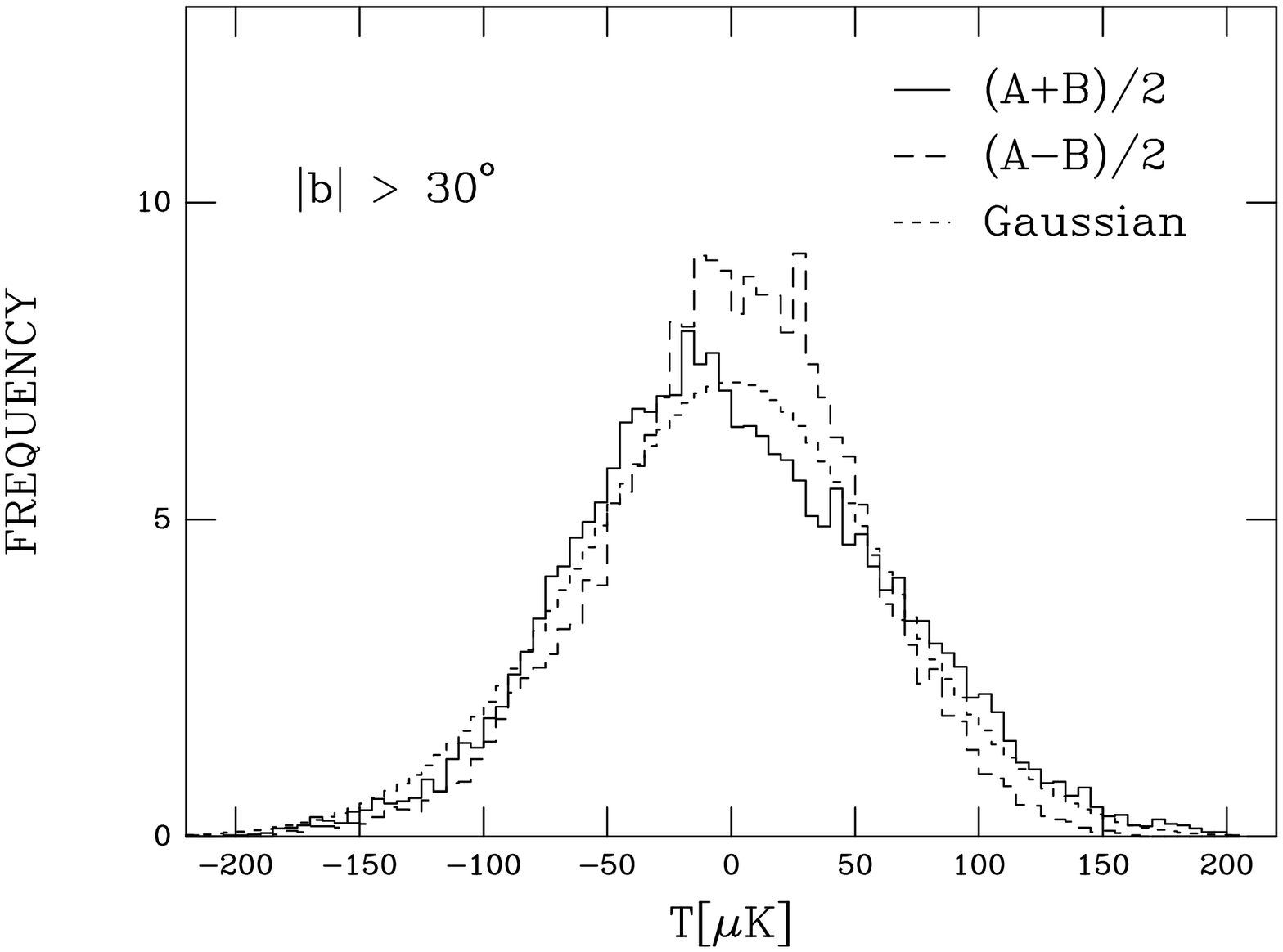}
\caption{The histogram of the smoothed (A-B)/2 test case map,
and its convolution by a Gaussian, which fails to match the observed 
histogram of the (A+B)/2 map.}
\label{testhist}
\end{figure}

The inflationary scenario also predicts that the temperature fluctuations
should have a Gaussian distribution.  This can be tested by looking at
higher-order moments of the maps such as three point correlation functions,
or by directly studying the probability distributions.  The random
measurement noise from the radiometers is still sufficient to interfere
with these studies to a great extent, but a preliminary analysis finds
that the Gaussian model is consistent with the observations.
Figure \ref{hist} shows histograms constructed from the smoothed DMR maps.
To minimize the radiometer noise, the thermodynamic $\Delta T$'s derived
from the 53 and 90 GHz channels have been averaged
together with weights of 0.6 and 0.4 respectively.
The histogram of the $(A-B)/2$ difference
map is not expected to be Gaussian, because of the large variation in
integration time per pixel seen in Figure \ref{DIAG}.  However, a Gaussian
convolved with the $(A-B)/2$ histogram should match the histogram of the
$(A+B)/2$ map if the sky has a Gaussian distribution.  As Figure \ref{hist}
shows, the ``Gaussian'' curve, which is this convolution, matches the
observed histogram well.
Detailed calculations show that models which predict
non-Gaussian temperature fluctuations, such as cosmic strings$^{20}$
or global monopoles$^{21}$, 
can also match the observed histogram.
This convergence toward Gaussian probability distributions is caused by
the 10\deg effective resolution of the smoothed maps, which averages
many small patches together.
Thus cosmic strings, which predict dramatically non-Gaussian behavior
(sharp edges) on small scales, give almost Gaussian results after
smoothing.
However, some models could be rejected: Figure \ref{testhist} shows the
histograms from the test case maps in Figure \ref{testmaps}.
Here the non-Gaussian structure is correlated over large distances, so
the smoothing does not erase it.

The ratio of the gravitational potential fluctuations seen by the DMR
through the Sachs-Wolfe$^{22}$ effect to the gravitational potentials
inferred from the bulk flows of galaxies using the POTENT method$^{23}$,
shows that the Harrison-Zeldovich spectrum
has the correct slope to connect the DMR data at $3\times 10^5$~km/sec
scales to the bulk flow data at 6000~km/sec scale.
This comparison of the horizon-scale perturbations seen by \COBE\ and
the ``large-scale'' ($60/h$~Mpc, where $h = H_\circ/(100\;\kmsMpc)$)
structure data gives best determination
of the slope of the power spectrum of the primordial perturbations,
even though it is model-dependent.  \COBE\ data alone cover such a
small range of scales that slope determinations are uncertain.

Wright \etal$^{16}$ discussed the implications of the \COBE\ DMR
data for models of structure formation, and selected 4 models
from a large collection$^{24}$ for detailed discussion:
a ``CDM'' model, with $H_\circ = 50\;\kmsMpc$, $\Omega_{CDM} = 0.9$,
and $\Omega_B = 0.1$;
a mixed ``CDM+HDM'' model, 
with $H_\circ = 50\;\kmsMpc$, $\Omega_{CDM} = 0.6$,
$\Omega_{HDM} = 0.3$ (a 7 eV neutrino), and $\Omega_B = 0.1$;
an open model, with $H_\circ = 100\;\kmsMpc$, $\Omega_{CDM} = 0.18$,
and $\Omega_B = 0.02$; and
a vacuum dominated model, 
with $H_\circ = 100\;\kmsMpc$, $\Omega_{CDM} = 0.18$,
$\Omega_B = 0.02$, and $\Omega_{vac} = 0.8$.
The vacuum dominated model and especially the open model
have potential perturbations now that are too small to explain the
POTENT bulk flows$^{23}$.
Note that all of these models are normalized to the \COBE\ anisotropy at
large scales at $z = 10^3$, but in the open and vacuum-dominated models
the potential perturbations go down even at the largest scales which
are not affected by non-linearities or the transfer function.
The small-scale anisotropies predicted by these models are close
to the upper limits, and in the case of the $1.5^\circ$ beam South Pole
experiment$^{25}$ (SP) they exceed the current data.  
Since the DMR datum is at $\ell_{eff} = 4$
while the SP datum is at $\ell_{eff} = 44$, a tilted model
which replaces the Harrison-Zeldovich $n = 1$ in $P(k) \propto k^n$
with an $n \approx 0.5$ would eliminate this discrepancy.  But such a low
$n$ would destroy the agreement between the bulk flow velocities and
the COBE \Amp.  
In fact the POTENT bulk flows are inconsistent with the SP
$\Delta T$ data$^{26}$, while other bulk flow determinations
are even larger$^{27}$.

\begin{figure}[tbp]
\plotone{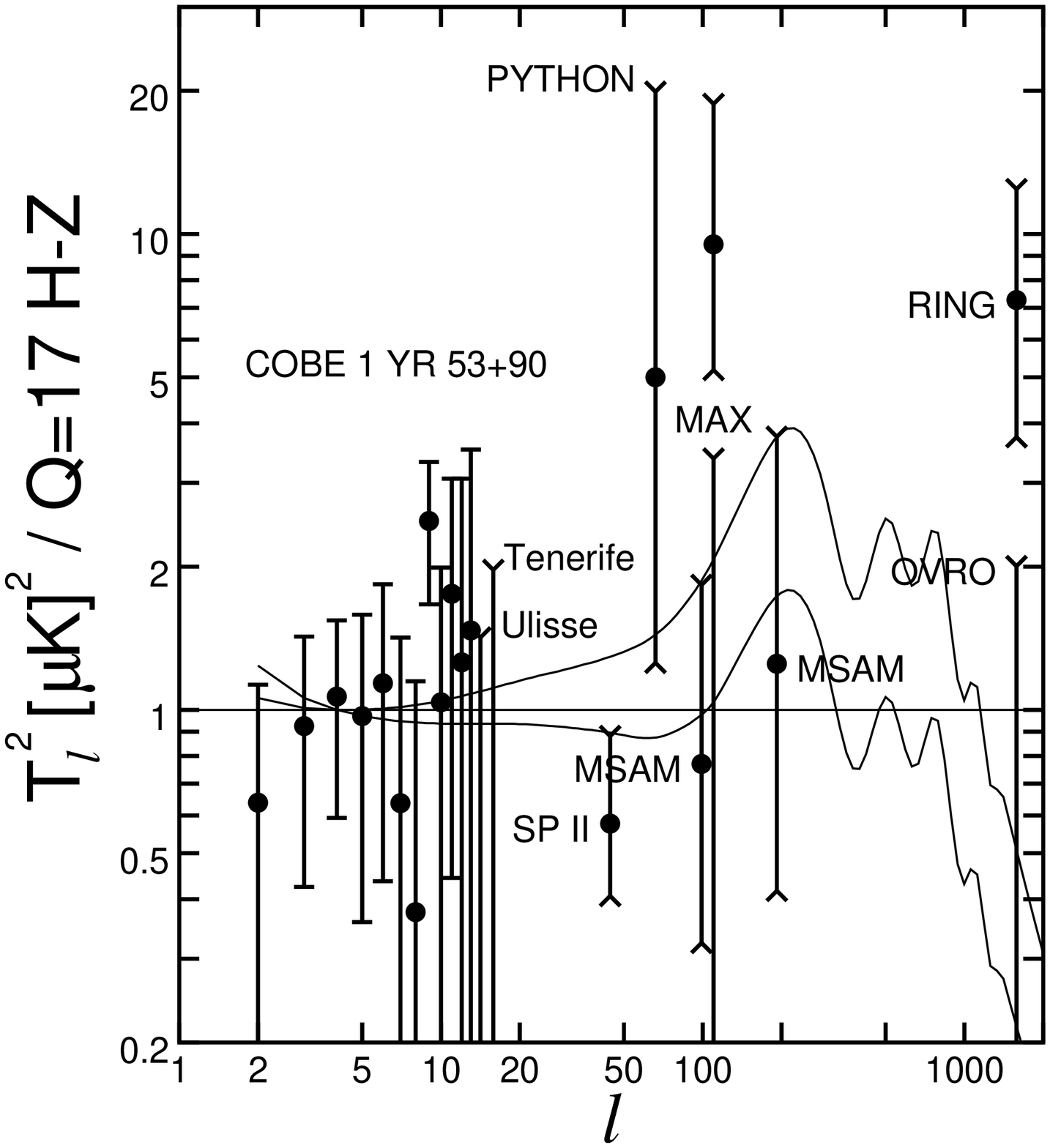}
\caption{
The angular power spectrum of the CMBR, normalized to an $\Amp = 17\;\muK$
Harrison-Zeldovich model.  Data from $\ell = 2..13$ are from COBE.
Other data and upper limits are from left to right:
Ulisse$^{32}$,
Tenerife$^{33}$,
SP II$^{25}$,
PYTHON$^{34}$,
MSAM$^{35}$,
two points from MAX$^{36,37}$,
MSAM again,
RING$^{38}$ and
OVRO$^{39}$.
}
\label{lg1yrmax}
\end{figure}

Wright$^{28}$ has computed the power spectrum of the DMR maps using
the Hauser-Peebles method to allow for incomplete sky coverage caused
by masking out the galactic plane region.
Figure \ref{lg1yrmax} shows the power in each multipole normalized to
a pure $n = 1$ (Harrison-Zeldovich) power law with $\Amp = 17\;\muK$.
Also shown are results from various small and medium scale $\Delta T$
experiments.
For perturbations at scales smaller than the horizon at the end of 
radiation dominance, dynamical effects boost the expected anisotropy.
At very small scales, the finite thickness of the recombination surface
filters out most of the anisotropy.
The models shown in Figure \ref{lg1yrmax}
are an $n = 0.85$ tilted CDM model$^{29}$ (lower curve, with 
$H_\circ = 50\;\kmsMpc$ and $\Omega_B h^2 = 0.0125$), and the same
model with the $n = 0.96$ expected from chaotic $\phi^4$ inflation.
The POTENT bulk flow, if translated into a 
$\Delta T_\ell^2$ in Figure \ref{lg1yrmax},
would correspond to a point on the higher model 
curve at $\ell \approx 100$.
The Lauer \& Postman$^{27}$ bulk flow would correspond to 
a $\Delta T_\ell^2$
that is $\approx 40$ times higher than the higher 
model at $\ell \approx 40$.
Clearly the tilt of the primordial power spectrum
cannot be deterined without much better measurements of small scale 
$\Delta T$ and bulk flow velocities.
Note that the RING experiment$^{38}$ is known to be contaminated by
discrete sources, but the power spectrum of uncorrelated point sources
is $\propto \ell^2$ in Figure \ref{lg1yrmax}, so this population
has a negligible effect on the data at lower $\ell$'s for sources which 
emit a spectrum similar to the spectrum of the Milky Way.
When comparing these data to the \COBE\ DMR data the corrections for
the galactic plane cut and the non-Gaussian shape of the DMR beam
should be used$^{30}$.

\section{FIRAS Measurements}

The Far InfraRed Absolute Spectrophotometer instrument on \COBE\ is
a polarizing Michelson$^{31}$ interferometer.  
The optical layout
is symmetrical, and it has two inputs and two outputs.  
If the two inputs
are denoted SKY and ICAL, then the two outputs, which are denoted LEFT and
RIGHT, are given symbolically as 
\begin{eqnarray}
{\rm LEFT} & = & {\rm SKY} - {\rm ICAL}\\
{\rm RIGHT} & = & {\rm ICAL} - {\rm SKY}
\end{eqnarray}
The FIRAS has achieved its incredible sensitivity to 
small deviations from a
blackbody spectrum by connecting the ICAL input to an 
internal calibrator, a
reference blackbody that can be set to a temperature close to the
temperature $T_\circ$ of the sky.  Thus this ``absolute'' 
spectrophotometer
is so successful because it is differential.
In addition, each output is further divided by a dichroic 
beamsplitter into a
low frequency channel (2-21 \percm) 
and a high frequency channel (23-95 \percm).
Thus there are four overall outputs.  These are labeled LL (left low)
through RH (right high).

Each output has a large composite
bolometer that senses the output power by absorbing the radiation, 
converting
this power into heat, and then detecting the temperature rise of a 
substrate
using a small and sensitive silicon resistance thermometer.  In order to
minimize the specific heat of the bolometer, which maximizes the 
temperature
rise for a given input power, the substrate is made of the material 
with the
highest known Debye temperature: diamond.  
Diamond is transparent to millimeter waves, however, so an absorbing
layer is needed.  Traditionally a layer of bismuth has been used as the
absorber on composite bolometers, but the FIRAS detectors use a thin layer
of gold alloyed with chromium.  
The absorbing layer needs to have a surface impedance about one-half of the
377 Ohms/$\Box$ impedance of free space in order to absorb efficiently, 
and this surface impedance can be obtained with a much thinner 
layer of gold rather than bismuth.  
Again, this serves to minimize the specific heat of the
detector.  In order to detect the longest waves
observed by FIRAS, the diameter of the octagonal diamond substrate is
large: about 8 mm.  Each detector sits behind a compound parabolic
concentrator, or Winston cone, which provides a full $\pi$ steradians 
in the
incoming beam.  As a result, the FIRAS instrument has a large \'etendue
of 1.5 cm$^2\;$sr.

\begin{figure}[tb]
\plotone{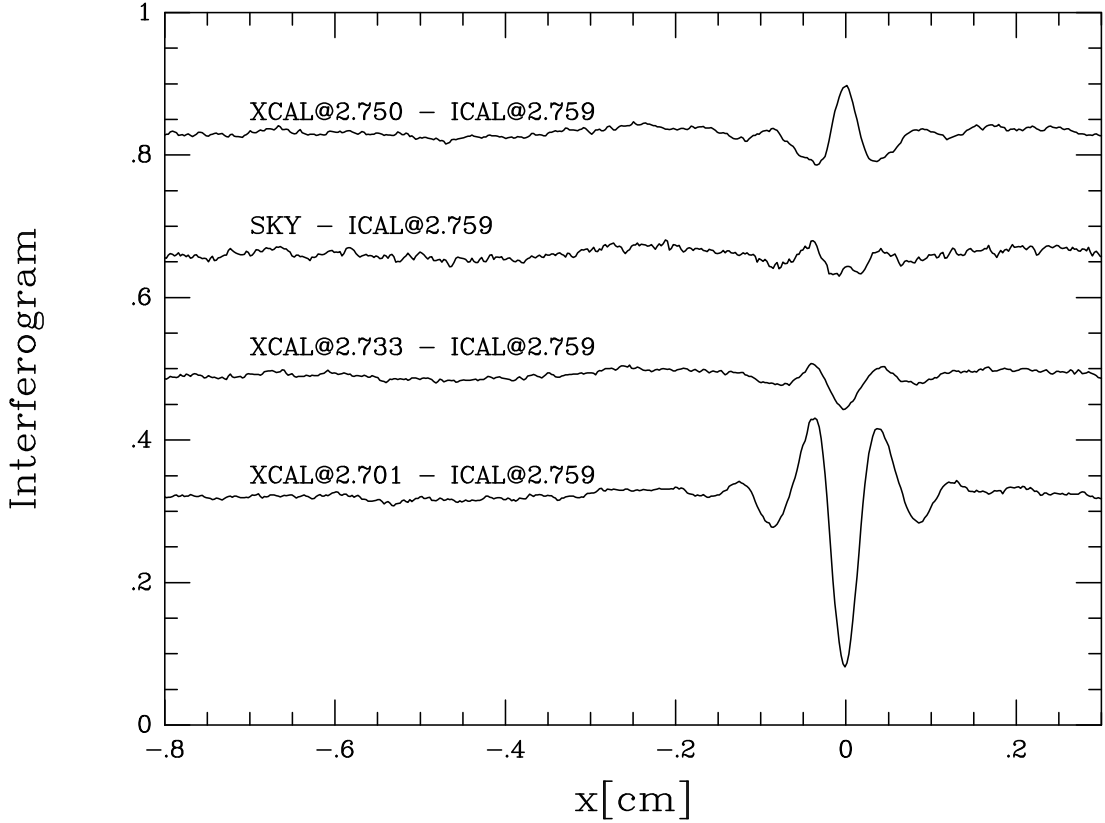}
\caption{Co-added FIRAS interferograms from the Left Low channel
taken during flight with the ICAL at 2.759~K while looking at the sky
or the XCAL at 2.701, 2.733 or 2.750~K.}
\label{ifg}
\end{figure}

Since the FIRAS is a Michelson interferometer, the spectral data are 
obtained
in the form of interferograms.  Thus the LEFT output is approximately
\begin{equation}
I_L(x) = \int_0^\infty \cos(2\pi\nu x) 
G(\nu) \left(I_\nu + \sum_i \epsilon_i(\nu) B_\nu(T_i) + U_\nu \right) d\nu
\end{equation}
The index $i$ above runs over all the components in the FIRAS that had
thermometers to measure $T_i$.  
These include the ICAL, the reference horn that
connects to the ICAL, the sky horn, the bolometer housing, the optical
structure of the FIRAS, and the dihedral mirrors that move to provide the
variation in path length difference $x$.
The $\epsilon_i$'s are the effective emissivities of the various 
components.
The ICAL itself has $\epsilon \approx -1$, while the sky and 
reference horns
have $\epsilon$'s of $\pm$ a few percent in the low frequency channels.
The offset term $U_\nu$ was observed during flight to be approximately
$10^{-5}\exp(-t/\tau)B_\nu(T_U)$ with a time constant $\tau$ of two months
and a temperature $T_U \approx 15$~K.
The SKY input $I_\nu$ above can be either the sky or an external 
calibrator.
The XCAL is a movable re-entrant absorber that can be inserted at the 
top of the sky horn.  The combination of sky horn plus XCAL forms a 
cavity with an absorbtivity known to be $> 0.9999$ from measurements,
and believed to be $\gsim 0.99999$ from calculations.
With the XCAL inserted during periodic calibration runs, the SKY input is
known to be $B_\nu(T_X)$.  By varying $T_X$ and the other $T_i$'s, the
calibration coefficients have been determined$^{40}$.  
Figure \ref{ifg}
shows the interferograms from flight used to determine the preliminary
FIRAS spectrum$^{41}$.

\begin{figure}[tb]
\plotone{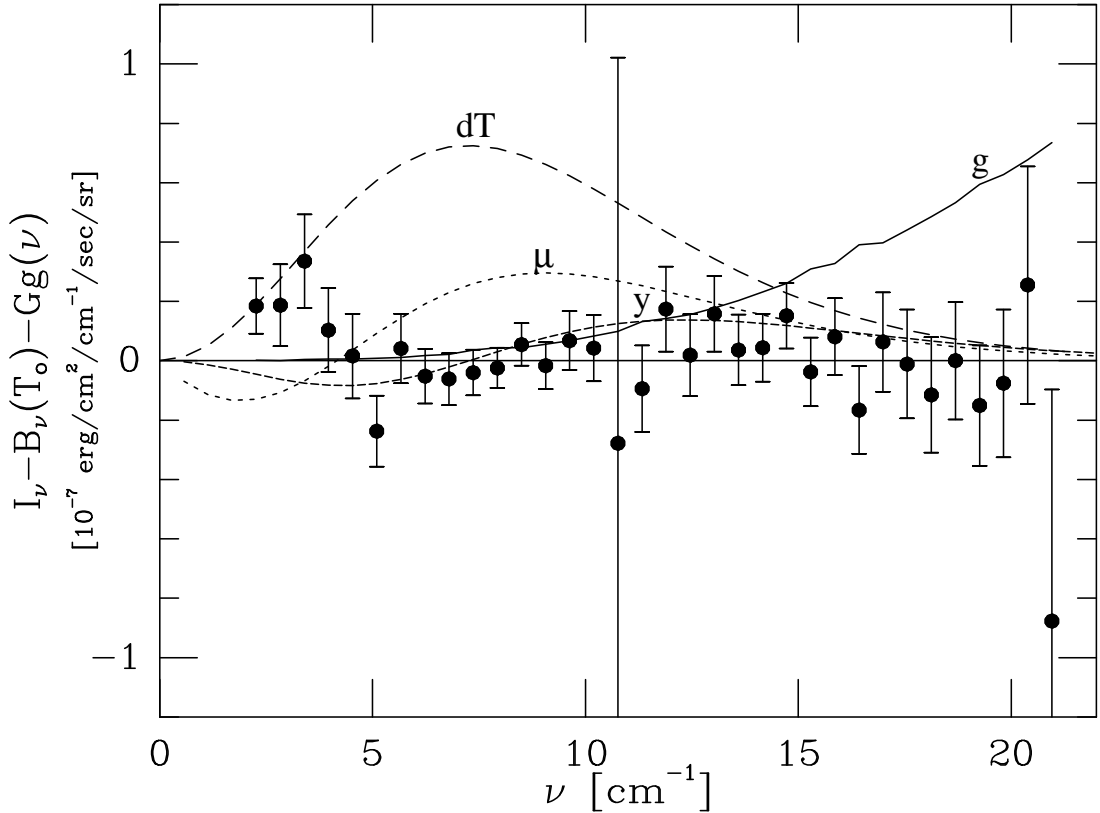}
\caption{FIRAS residuals (points) after fitting for a blackbody and
galactic emission, with an 0.5~mK $\Delta T$, a
$\Delta(\csc|b|) = 0.25$ galaxy curve, $y = 2.5\times 10^{-5}$,
and $\mu = 3.3\times 10^{-4}$ shown for comparison.  Full-scale
is 0.1\% of the peak of the CMB.}
\label{rtg}
\end{figure}

\begin{figure}[tb]
\plotone{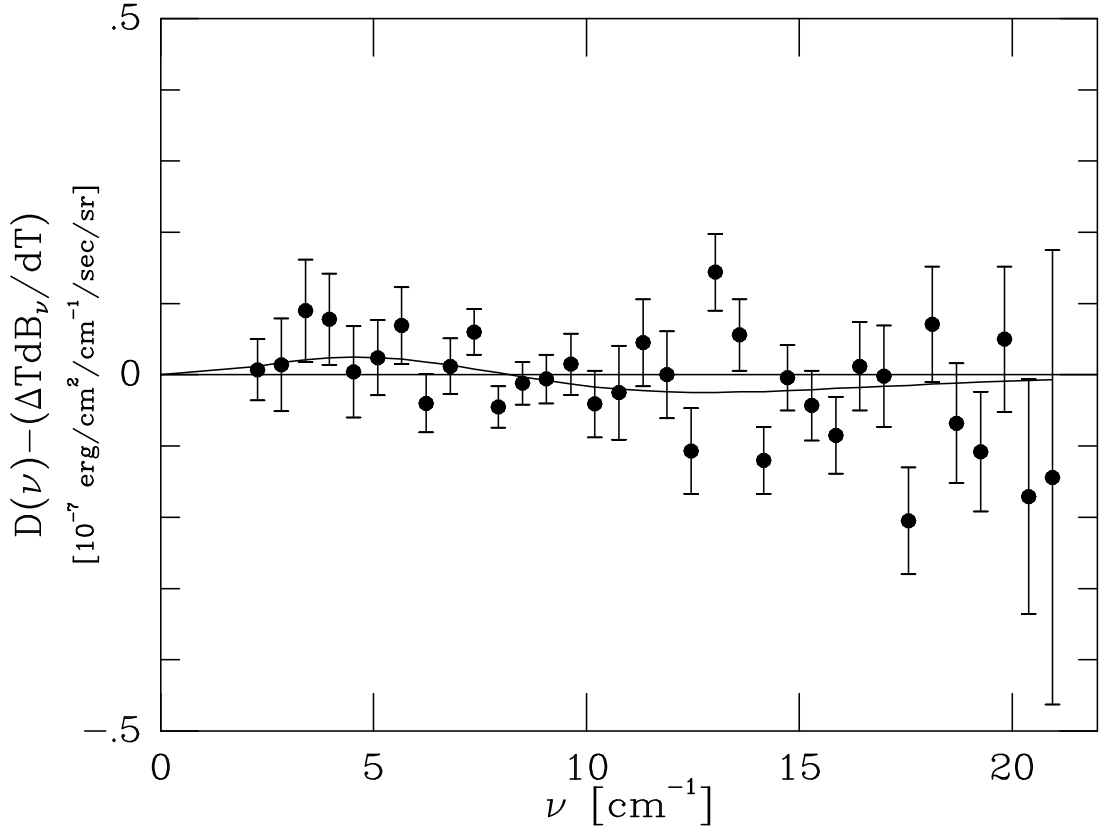}
\caption{FIRAS Dipole spectrum residuals from an amplitude of
3.343 mK applied to $T_\circ = 2.726$~K (points).
Full-scale is 0.04\% of the peak of the CMB.
The curve shows the best fit dipole color temperature of 2.714 K
with an amplitude of 3.379 mK.}
\label{dipres}
\end{figure}

Once the calibration coefficients are known, the sky data can be 
analyzed to
determine $I_\nu(l,b)$ the intensity of the sky as a function of 
frequency, galactic longitude and galactic latitude.  
This is a ``data cube''.  
The data from each direction on the sky can be written as a 
combination of cosmic plus galactic signals:
\begin{eqnarray}
I_\nu(l,b) & = & e^{-\tau_\nu(l,b,\infty)} 
\left(B_\nu\left(T_\circ + \Delta T(l,b)\right)
 + \Delta I_\nu \right)\nonumber \\
& + & \int e^{-\tau_\nu(l,b,s)} j_\nu(l,b,s) ds
\end{eqnarray}
where $\tau_\nu(l,b,s)$ is the optical depth between the Solar system and 
the point at distance $s$ in the direction $(l,b)$ at frequency $\nu$,
$\Delta I_\nu$ is an isotropic cosmic distortion, and
$\Delta T(l,b)$ is the variation of 
the background temperature around its mean
value $T_\circ$.  This equation can be simplified because the optical
depth of the galactic dust emission is always small in the
millimeter and sub-millimeter bands covered by FIRAS.  
\begin{equation}
I_\nu(l,b)  \approx 
B_\nu\left(T_\circ + \Delta T(l,b)\right) + \Delta I_\nu
 +  \int j_\nu(l,b,s) ds
\end{equation}
Even in the optically
thin limit, some restrictive assumptions about the galactic emissivity
$j_\nu$ are needed, since the galactic intensity
$\int j_\nu(l,b,s) ds$ is a function of three variables, just like the
observed data.  The simplest reasonable model$^{42}$ for the 
galactic emission
is
\begin{equation}
\int j_\nu(l,b,s) ds = G(l,b) g(\nu). 
\end{equation}
This model assumes that the shape of the galactic spectrum is independent
of direction on the sky.  It is reasonably successful except that the 
galactic center region is clearly hotter than the rest of the galaxy.
The application of this model proceeds in two steps.  The first step
assumes that the cosmic distortions vanish, and that an approximation
$g_\circ(\nu)$ to the galactic spectrum is known.  A least squares fit
over the spectrum in each pixel then gives the maps $\Delta T(l,b)$
and $G(l,b)$.  The high frequency channel of FIRAS is used to derive
$G(l,b)$ because the galactic emission is strongest there.  An alternative
way to derive $G(l,b)$ is to smooth the DIRBE map at 240 \um\ to the FIRAS
$7^\circ$ beam.  This DIRBE method has been used in the latest 
FIRAS spectral results$^{43,44,45}$.
The second step in the galactic fitting then derives
spectra associated with the main components of the millimeter wave sky:
the isotropic cosmic background, the dipole anisotropy, and the galactic
emission.  This fit is done by fitting all the pixels (except for 
the galactic center region with $|b| < 20$ and $|l| < 40$)
at each frequency to the form
\begin{equation}
I_\nu(l,b) = I_\circ(\nu) + D(\nu)\cos\theta + G(l,b)g(\nu).
\end{equation}
The spectra of the anisotropic components derived from this fit applied
to the entire mission data set are shown in Figures \ref{rtg} and
\ref{dipres}.  Figure \ref{dipres} shows the residual$^{44}$ after the
predicted dipole spectrum $\Delta T \partial B_\nu/\partial T$ evaluated
at $T_\circ$ is subtracted from $D(\nu)$.

But the FIRAS calibration model is so complicated and the cosmic 
distortions,
if any, are so small, that there are still systematic uncertainties that
limit the accuracy of the isotropic spectrum $I_\circ(\nu)$ 
derived from the
whole FIRAS data set.  Therefore the best estimate of $I_\circ(\nu)$ comes
from the last six weeks of the mission, when a modified observing sequence
consisting of 3.5 days of observing the sky with the ICAL set to null out
the cosmic spectrum, followed by 3.5 days of observing the XCAL with its
temperature set to match the sky temperature $T_\circ$.  Six cycles of
this alternation between sky and XCAL were obtained before the helium
ran out.  By processing both the sky data and the XCAL data through the
same calibration model, and then subtracting the two spectra,
almost all of the systematic errors cancel out.  The residual in the
high galactic latitude sky is then modeled as
\begin{equation}
I_\nu(sky,|b| > b_c) - I_\nu(XCAL) = \Delta I_\nu + 
\delta T \frac {\partial B_\nu} {\partial T} + G g(\nu).
\label{resid}
\end{equation}
The result of a least squares fit to minimize $\Delta I_\nu$ by
adjusting $\delta T$ and $G$ is shown in Figure \ref{rtg}.
Since all of the data taken during the last six weeks of the mission
were taken in a single scan mode, the mechanical resonance frequency
of the mirror carriage occurs at a fixed spectral frequency, leading 
to the large error bar on the point at 11 \percm.
The data points show the residuals while the curves show the
``uninteresting'' models $\partial B_\nu/\partial T$ and $g(\nu)$,
and the ``interesting'' $y$ and $\mu$ models.
It is important to remember that any cosmic distortion with a spectral
shape that matches one of the uninteresting curves will be hidden in this 
fit.
The maximum residual between 2 and 20 \percm\ is 1 part in 3000 of the peak
of the blackbody, and the weighted RMS residual is 1 part in $10^4$ of the
peak of the blackbody.
While these residuals are small, they are nonetheless more than twice
the residuals that one would expect from detector noise alone.  The error
bars in Figure \ref{rtg} have been inflated by a constant factor to
give $\chi^2 = 32$ for the 32 degrees of freedom in the fit.
Limits on cosmological distortions can now be set by adding terms to the
above fit, finding the best fit value with minimum $\chi^2$, and then
finding the endpoints of the 95\% confidence interval where 
$\chi^2 = \chi^2_{min} + 4$.  
As an example of such a fit, suppose that the sky in fact was a gray-body
with an emissivity $\epsilon = 1+e$, where $e$ is a small parameter.
Since the XCAL is a good blackbody, this model predicts a residual of the
form $eB_\nu(T_\circ)$.  
Because $B_\nu(T_\circ)$ peaks at a lower frequency than
$\partial B_\nu/\partial T$, this model is sufficiently different from the
``uninteresting'' models and a tight limit on $e$ can be set:
$|e| < 0.00041$ (95\% confidence).
The result of fits for the usual cosmological suspects are
$|y| < 2.5 \times 10^{-5}$ and $|\mu| < 3.3 \times 10^{-4}$ with 95\%
confidence.

The absolute temperature of the cosmic background, $T_\circ$, can be
determined two ways using FIRAS.  The first way is to use the readings
of the germanium resistance thermometers in the XCAL when the XCAL
temperature is set to match the sky.  This gives $T_\circ = 2.730$~K.
The second way is to measure the frequency of the peak of 
$\partial B_\nu/\partial T$ by varying $T_X$ a small amount
around the temperature which matches the sky, and then apply
the Wien displacement law to convert this frequency into a
temperature.  This calculation is done automatically by the
calibration software, and it gives a value of 2.722~K for
$T_\circ$.  
Three additional determinations of $T_\circ$ depend on the dipole
anisotropy.
For either FIRAS or DMR, the spectrum of the dipole anisotropy
can be fit to the form
\begin{equation}
D(\nu) = \frac{T_\circ v}{c} \frac {\partial B_\nu(T_\circ)} {\partial T}
\end{equation}
Since the velocity of the solar system with respect to the CMB is not known
{\it a priori}, only the shape and not the amplitude of the dipole spectrum
can be used to determine $T_\circ$.
For FIRAS, this analysis gives $T_\circ = 2.714 \pm 0.022$~K$^{44}$,
while for DMR it gives $T_\circ = 2.76 \pm 0.18$~K$^{12}$.
The DMR data analysis keeps track of the changes in the dipole caused
by the variation of the Earth's velocity around the Sun during the year.
In this case the velocity $v$ is known, so $T_\circ$ can be determined
from the amplitude of the change in the dipole, giving 
$T_\circ = 2.75 \pm 0.05$~K$^{12}$.
The final adopted value is $2.726 \pm 0.010$~K$^{43}$
(95\% confidence),
which just splits the difference between the two methods based on the
FIRAS spectra.
The dipole-based determinations of $T_\circ$ are less precise but provide
a useful confirmation of the spectral data.

\section{FIRAS Interpretation}

When interpreting the implications of the FIRAS spectrum on effects
that could distort the microwave background, it is important to
remember some basic order-of-magnitude facts.  Theoretical analyses
of light element abundances gives limits on the current density of
baryons through models of Big Bang NucleoSynthesis (BBNS).
The ratio of baryons to photons$^{46}$ is
$\eta = (3.5 \pm 0.7) \times 10^{-10}$, while another determination$^{47}$
gives $\eta = (3.95 \pm 0.25) \times 10^{-10}$.  
As a result, the effect of Lyman $\alpha$ photons from recombination
is negligible, simply because there are so few hydrogen atoms per photon.
Since the number
density of photons is well determined by the FIRAS spectrum
\begin{equation}
N = 8 \pi \zeta(3)\Gamma(3) \left( \frac {k T_\circ} {h c} \right)^3
  = 410 \pm 5\;\percmcub
\end{equation}
the baryon density is well determined: $1.62 \times 10^{-7}\;\percmcub$.
This gives $\Omega_B h^2 = 0.0144$.
Not all of these baryons are protons, however.  For a primordial
helium abundance of 23.5\% by mass, one has 13 protons and 1 $\alpha$
particle in 17 baryons.  This gives an electron density of 
$n_{e,\circ} = 1.43 \times 10^{-7}\;\percmcub$.
For a fully ionized Universe, the optical depth for electron scattering
over a path length equal to the Hubble radius $c/H_\circ$ is $0.00088/h$.

The expansion of the Universe $H$ varies with time and redshift in a way
that depends on ratios of the current matter, radiation and 
vacuum densities
to the critical density $\rho_c = 3H^2/8\pi G$.  Let these ratios be
$\Omega_m$, $\Omega_r$, and $\Omega_v$.  Then
\begin{equation}
H(z) = (1+z) H_\circ \sqrt{(1-\Omega_m-\Omega_r-\Omega_v) + \Omega_m(1+z)
+ \Omega_r (1+z)^2 + \Omega_v(1+z)^{-2}}
\end{equation}
This equation is not exact for massive neutrinos which shift from being
part of $\Omega_m$ into part of $\Omega_r$ as $z$ increases.
The value of $\Omega_r h^2 = (4.16 \pm 0.06)\times 10^{-5}$ 
is fairly well-determined for the case of three massless neutrinos.
Thus if $\Omega_m \approx 1$ now,
the Universe becomes radiation dominated for 
$z > z_{eq} = 2.4h^2 \times 10^4$ when the background temperature was
$kT_{eq} = 5.65h^2\;{\rm eV}$.
In the radiation dominated epoch, $H = 0.645 (1+z)^2\;{\rm km/sec/Mpc}$ or
$H = 2.09(1+z)^2 \times 10^{-20}\;{\rm s}^{-1}$.

For early epochs when the Universe was ionized, electron scattering is
the dominant mechanism for transferring energy between the radiation field
and the matter.  As long as the electron temperature is less than
about $10^8$~K, the effect of electron scattering on the spectrum can
be calculated using the Kompaneets equation$^{48}$:
\begin{equation}
\frac {\partial n}{\partial y} = x^{-2} \frac{\partial}{\partial x}
\left[x^4\left(n + n^2 + \frac{\partial n}{\partial x}\right)\right]
\end{equation}
where $n$ is the number of photons 
per mode ($n = 1/(e^x-1)$ for a blackbody),
$x = h\nu/kT_e$, and the Kompaneets $y$ is defined by
\begin{equation}
dy = \frac{k T_e}{m_e c^2} n_e \sigma_T c dt.
\end{equation}
Thus $y$ is the electron scattering optical depth times the electron 
temperature in units of the electron rest mass.  Note that the electrons
are assumed to follow a Maxwellian distribution, but that the photon
spectrum is completely arbitrary.

Since the Kompaneets equation is describing electron scattering, which
preserves the number of photons, one finds that the $y$ derivative of the
photon density $N$ vanishes:
\begin{eqnarray}
\frac {\partial N}{\partial y} & \propto &\int x^2 \frac {\partial n}
{\partial y} dx \nonumber \\
 & = & \int \frac{\partial}{\partial x}
\left[x^4\left(n + n^2 + \frac{\partial n}{\partial x}\right)\right] dx 
\nonumber \\
 & = & 0 
\end{eqnarray}
The stationary solutions of the equation 
$\partial n/\partial y = 0$
are the photon distributions in thermal equilibrium with the electrons.
Since photons are conserved, the photon number density does not have to
agree with the photon number density in a blackbody at the electron
temperature.  Thus a more general Bose-Einstein thermal distribution
is allowed: $n = 1/(\exp(x+\mu)-1)$.  This gives
$\partial n/\partial y = 0$ for all $\mu$.
Since the Bose-Einstein spectrum is a stationary point of the Kompaneets
equation, it is the expected form for distortions produced at epochs
when
\begin{equation}
(1+z)\frac{\partial y}{\partial z} = 
\sigma_T n_{e,\circ} \frac {k T_\circ}{m_e c^2} \frac {c}{H}(1+z)^4 > 1
\end{equation}
For the $\Omega_B h^2$ given by BBNS, this redshift $z_y$ where
this inequality is crossed is well within the 
radiation dominated era, with a value
$z_y = 10^{5.1}/\sqrt{70\Omega_B h^2}$.

There is a simple solution to the Kompaneets equation with non-zero
$\partial n/\partial y$ which gives the 
Sunyaev-Zeldovich$^{49}$ or $y$ spectral distortion.
This simple case occurs when the initial photon field is a blackbody
with a temperature $T_\gamma$ which is below the electron 
temperature.  Letting $f = T_e/T_\gamma$, we find that the initial photon
field is given by $n = 1/(\exp(fx)-1)$.  Therefore
\begin{equation}
\left(n + n^2 + \frac{\partial n}{\partial x}\right) = 
\frac{(1-f)\exp(fx)}{(\exp(fx)-1)^2} 
= (1-f^{-1}) \frac{\partial n} {\partial x}.
\end{equation}
Thus the distortion has a Sunyaev-Zeldovich shape but is reduced in
magnitude by a factor $(1-T_\gamma/T_e)$.  Defining the ``distorting''
$y$ as
\begin{equation}
dy_D = \frac{k (T_e - T_\gamma)}{m_e c^2} n_e \sigma_T c dt
\end{equation}
we find that the final spectrum is given by a frequency-dependent
temperature given by
\begin{equation}
T_\nu = T_\circ \left[1 + y_D
\left(\frac {x (e^x+1)}{e^x-1} - 4 \right) + \ldots \right].
\end{equation}
where $x = h\nu/kT_\circ$.
The FIRAS spectrum in Figure \ref{rtg} shows that 
$|y_D| < 2.5 \times 10^{-5}$.

\begin{figure}[tb]
\plotone{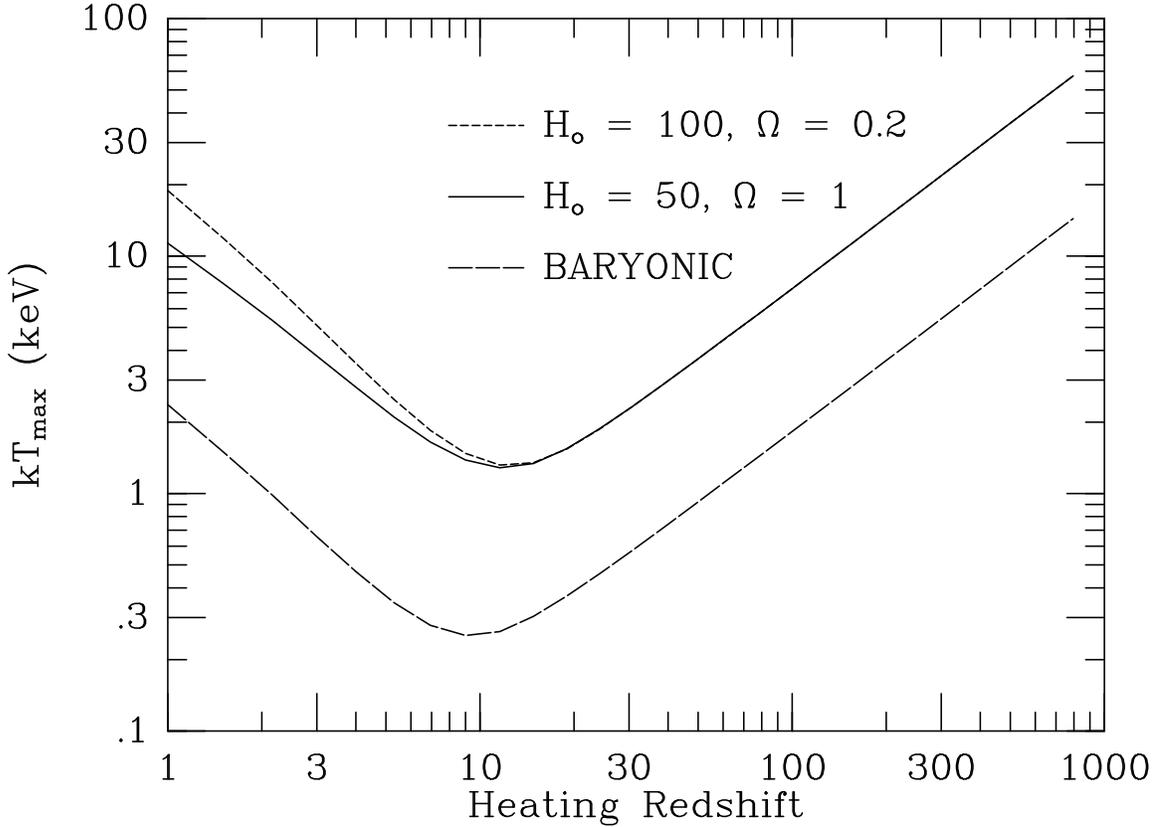}
\caption{The maximum allowed temperature for an IGM as a function of the
heating redshift for two models with the BBNS baryon abundance 
($\Omega_B h^2 = 0.0125$) and
one model with 4 times more baryons ($H_\circ = 50, \Omega = \Omega_B =
0.2$).}
\label{tmax}
\end{figure}

The energy density transferred from the hotter electrons to the cooler
photons in the $y$ distortion is easily computed.  The energy density is
given by $U \propto \int x^3 n dx$ so
\begin{equation}
\frac {\partial U}{\partial y_D} = 
\int x \frac {\partial}{\partial x}\left(x^4 \frac{\partial n}{\partial x}
\right) dx
\end{equation}
which when integrated by parts twice gives
\begin{eqnarray}
\frac {\partial U}{\partial y_D} & = &
-\int \left(x^4 \frac{\partial n}{\partial x} \right) dx \nonumber \\
 & = & 4 \int x^3 n dx = 4 U.
\end{eqnarray}
Thus the limit on $y_D$ gives a corresponding limit on energy transfer:
$\Delta U/U < 10^{-4}$.
Any energy which is transferred into the electrons at redshifts $z > 7$
where the Compton cooling time is less than the Hubble time will be
transferred into the photon field and produce a $y$ distortion.
Since there are $10^9$ times more photons than any other particles except
for the neutrinos, the specific heat of the photon gas is overwhelmingly 
dominant, and the electrons rapidly cool (in a Compton cooling time) back
into equilibrium with the photons.  The energy gained by the photons
is $\Delta U = 4 y a T_\gamma^4$ which must be equal to the energy
lost by the electrons and ions: $1.5 (n_e + n_i) k \Delta T_e$.  Since 
$y = \sigma_T n_e (k T_e/m_e c^2) c \Delta t$ we find the Compton cooling
time
\begin{equation}
t_C = \frac{1.5 (1+n_i/n_e) m_e c^2 } {4\sigma_T c U_{rad}}
    = \frac{7.4 \times 10^{19}\;{\rm sec}}{(1+z)^4}
\end{equation}
for an ion to electron ratio of $14/15$.
This becomes equal to the Hubble time 
$(3.08568\times 10^{17} {\rm sec})/(h(1+z)^{1.5})$ at $(1+z) = 9h^{0.4}$.
Figure \ref{tmax} shows the maximum allowed temperature$^{45}$ for an
intergalactic medium heated impulsively as a function of the heating
redshift.  After being heated, the IGM cools by adiabatic expansion and
Compton cooling.  The 3 models shown in Figure \ref{tmax} are an open
model with $H_\circ = 100 \; \kmsMpc$, $\Omega_B h^2 = 0.0125$, and
$\Omega = 0.2$; a flat model with $H_\circ = 50$, $\Omega = 1$ and the
same $\Omega_B$; and a ``BARYONIC'' model with $H_\circ = 50$ and
$\Omega = \Omega_B = 0.2$ which has four times more baryons than BBNS
models predict.  For this increased value of the baryon abundance, the
allowed $kT_{max}$ is considerably less than the 1 keV suggested by the
300 km/sec velocity dispersion in galaxies.

At redshifts $z > z_y = 10^5$, there will be enough electron scattering to
force the photons into a thermal distribution with a $\mu$ distortion
instead of a $y$ distortion.  However, the normal form for writing a $\mu$
distortion does not preserve the photon number density.  Thus we should 
combine the $\mu$ distortion with a temperature change to give an effect
that preserves photon number.  The photon number density change with $\mu$
is given by
\begin{eqnarray}
N  & \propto & \int {\frac {x^2 dx} {\exp(x+\mu)-1}} \nonumber \\
   & = & \sum_{k=1}^\infty e^{-k\mu} \int x^2 e^{-kx} dx \nonumber \\
   & = & 2 \sum_{k=1}^\infty {\frac {e^{-k\mu}}{k^3} } \nonumber \\
   & = & 2 \left(\zeta(3) - \mu \zeta(2) + \ldots \right).
\end{eqnarray}
A similar calculation for the energy density shows that
\begin{equation}
U \propto 6 \left(\zeta(4) - \mu \zeta(3) + \ldots \right).
\end{equation}
In order to maintain $N = const$, the temperature of the photon field
changes by an amount $\Delta T/T = \mu\zeta(2)/(3\zeta(3))$.
Therefore, the energy density change at constant $N$ is
\begin{equation}
\frac {\Delta U}{U} = 
\left(\frac {4\zeta(2)}{3\zeta(3)}-\frac{\zeta(3)}{\zeta(4)}\right) \mu
= 0.714\mu.
\end{equation}
Thus the FIRAS limit $|\mu| < 3.3 \times 10^{-4}$ implies 
$\Delta U/U < 2.4 \times 10^{-4}$.
The ``improved'' form of the $\mu$ distortion, with a $\Delta T$
added to keep $N$ constant, can be given as a frequency dependent 
brightness temperature:
\begin{equation}
T_\nu = T_\circ \left(1 + \mu 
\left[\frac{\zeta(2)}{3\zeta(3)} - x^{-1}\right] + \ldots \right).
\end{equation}
In this form it is clear that a $\mu$ distortion has a deficit of
low energy photons and a surplus of high energy photons with respect
to a blackbody.  In this it is like the $y$ distortion, but the
crossover frequency is lower.
The ``improved'' form of the $\mu$ distortion is plotted in 
Figure \ref{rtg}.

Finally, at high enough redshift the process of double photon Compton
scattering becomes fast enough to produce the extra photons needed to
convert a distorted spectrum into a blackbody.  Whenever a photon with
frequency $\nu$ scatters off an electron, there is an impulse 
$\propto h\nu/c$
transferred to the electron.  This corresponds to an acceleration
$a \propto h\nu^2/(m_ec)$ for a time interval $\Delta t \propto 1/\nu$.
The energy radiated in new photons is thus $\propto e^2h^2\nu^3/(m_e^2c^5)$
which is $\propto \alpha h\nu (h\nu/(m_ec^2))^2$.  Since the rate 
of scatterings per photon is $\propto \Omega_B h^2 (1+z)^3$, 
the overall rate of new photon creation is $\propto \Omega_B h^2 (1+z)^5$ 
while the Hubble time is $\propto (1+z)^{-2}$.
Thus the photon creation rate per Hubble time is $\propto (1+z)^3$
and at earlier enough times a blackbody spectrum is produced
no matter how much energy is transferred to the photon field.
The emissivity for double photon Compton scattering has approximately the 
same spectral shape as free-free emission, since both are due to
impulsive accelerations.
Thus the addition of photons to the radiation field can be described by
\begin{equation}
\frac {\partial n}{\partial y} = A \frac {1-e^{-x}}{x^3} 
\left( \frac {1}{e^x-1} - n \right)
\end{equation}
where $A$ is the ratio of photon creation via
double photon Compton scattering or free-free emission 
to the increase of $y$.
For double photon Compton scattering $A$ scales like $(1+z)$.
Because the photons are created primarily at low frequencies,
$n(x)$ approaches a blackbody for $x \rightarrow 0$, while the frequency
shifts due to Compton scattering are simultaneously transferring photons
to larger values of $x$.
$A$ is very small at redshifts where distortions can survive,
so the photon creation occurs primarily at $x \ll 1$.
For any $n = (\exp(x+\mu(x))-1)^{-1}$,
\begin{equation}
\left(n + n^2 + \frac {\partial n} {\partial x} \right) =
-n(n+1)\frac{\partial \mu}{\partial x}
\end{equation}
so when $\mu(x) = \mu\exp(-x_\circ/x)$
and $x, x_\circ, \mu$ and $A$ are all $\ll 1$, implying
$n \approx n+1 \approx x^{-1}$,
\begin{equation}
x^{-2} \frac{\partial}{\partial x}
\left( x^4 \left(n + n^2 + \frac {\partial n} {\partial x} \right) \right)
\approx \frac {-\mu \exp(-x_\circ/x) x_\circ^2} {x^4}
\end{equation}
This cancels the $\partial n/\partial y$ due to photon addition, giving
a quasi-equilibrium solution, if $x_\circ = \sqrt{A}$.
With this form for $n$ one finds a net photon addition rate of
\begin{eqnarray}
\frac{\partial N}{\partial y} & = & 
\int x^2 \frac {\partial n}{\partial y} dx
\nonumber \\
& = & \int x^2 A \frac {1-e^{-x}}{x^3} 
\left( \frac {1}{e^x-1} - n \right) dx
\nonumber \\
& = & A\mu/x_\circ = \mu \sqrt{A}.
\end{eqnarray}
Since the deficit of photons associated with $\mu$
is $\mu\pi^2/3$, one finds a thermalization rate per unit $y$ of
\begin{equation}
\frac {\partial \ln \mu}{\partial y} = \frac{3\sqrt{A}}{\pi^2}
\propto \sqrt{1+z}
\end{equation}
Since $(1+z)\partial y/\partial z \propto \Omega_B h^2 (1+z)^2$,
the overall rate for eliminating a $\mu$ distortion scales like
$\Omega_B h^2 (1+z)^{5/2}$ per Hubble time.
A proper consideration$^{50}$ of this interaction
of the photon creation process with the Kompaneets equation
shows that the redshift from which
$1/e$ of an initial distortion can survive is
\begin{equation}
z_{th} = \frac {4.24 \times 10^5}
{\left[\Omega_B h^2 \right]^{0.4}}
\end{equation}
which is $z_{th} = 2.3 \times 10^6$ for the BBNS value of $\Omega_B h^2$.

\section{Summary}

So far \COBE\ has been a remarkably successful space experiment with
dramatic observational consequences for cosmology, and the
DIRBE determination of the cosmic infrared background is yet to come.
The very tight limits on deviations of the spectrum from a blackbody
rule out many non-gravitational models for structure formation,
while the amplitude of the $\Delta T$ discovered by the \COBE\ DMR
implies a magnitude of gravitational forces in the Universe that is
sufficient to produce the observed clustering of galaxies,
{\it but only if the Universe is dominated by dark matter.}   

\section{References}

\setlength{\parskip}{2pt}

\refitem
1! 
Penzias, A. A. \& Wilson, R. W. 1965! ApJ! 142! 419;

\refitem
2! 
Boggess, N. W. \etal\ 1992! ApJ! 397! 420;

\bookref
3!
Bennett, C. L. \etal\ 1992a.  COBE Preprint 92-08, 
``Recent Results
from COBE'', to be published in The Third Teton Summer School: The
Evolution of Galaxies and Their Environment, eds. H. A Thronson \&
J. M. Shull.

\refitem
4!
Hubble, E. 1929! PNAS! 15! 168;

\bookref
5!
Peebles, P. J. E. 1993, ``Principles of Physical Cosmology'',
Princeton University Press.

\refitem
6!
Guth, A. 1981! Phys. Rev. D! 23! 347;

\refitem
7!
Partridge, R. B. \& Peebles, P. J. E. 1967! ApJ! 148! 377;

\refitem
8!
Stecker, F. W., Puget, J. L. \& Fazio, G. G. 1977! ApJL! 214! L51;

\refitem
9!
Dermott, S. F., Nicholson, P. D., Burns, J. A. \& Houck, J. R. 1984!
Nature! 312! 505-509;

\refitem
10!
Smoot, G. F. \etal\ 1992! ApJL! 396! L1;

\refitem
11!
Conklin, E. K. 1969! Nature! 222! 971;

\refitem
12!
Kogut, A., Lineweaver, C., Smoot, G.~F., Bennett, C.~L., Banday, A., 
Boggess, N.~W., 
Cheng, E.~S., De~Amici, G., Fixsen, D.~J., Hinshaw, G., Jackson, P.~D.,
Janssen, M., Keegstra, P., Loewenstein, K., Lubin, P., Mather, J.~C., 
Tenorio, L., Weiss, R., Wilkinson, D.~T., \& Wright, E.~L., 
1993! ApJ! 419! TBD;

\refitem
13!
Harrison, E. R. 1970! Phys Rev D! 1! 2726-2730;

\refitem
14!
Zeldovich, Ya. B. 1972! MNRAS! 160! 1p;

\refitem
15!
Peebles, P. J. E. \& Yu, J. T. 1970! ApJ! 162! 815-836;

\refitem
16!
Wright, E. L. \etal\ 1992! ApJL! 396! L13;

\refitem
17! Bennett, C. L. \etal\ 1992b! ApJL! 396! L7;

\refitem
18!
Kogut, A. \etal\ 1992! ApJ! 401! 1-18;

\refitem
19! 
Bennett, C. L., Hinshaw, G., Banday, A., Kogut, A., Wright, E. L.,
Loewenstein, K. \& Cheng, E. S. 1993! ApJL! 414! TBD;

\refitem
20! Bennett, D. P., Stebbins, A. \& Bouchet, F. R. 1992! ApJL! 399! L5-L8;

\refitem
21!
Bennett, D. P. \& Rhie, S. H. 1993! ApJ! 406! L7;

\refitem
22!
Sachs, R. K. \& Wolfe, A. M. 1967! ApJ! 147! 73;

\refitem
23!
Bertschinger, E., Dekel, A., Faber, S. M., Dressler, A. \& 
Burstein, D. 1990!
ApJ! 364! 370-395;

\refitem
24!
Holtzman, Jon A. 1989! ApJSupp! 71! 1-24;

\refitem
25!
Schuster, J., Gaier, T., Gundersen, J., Meinhold, P., Koch, T., 
Seiffert, M.,
Wuensche, C. \& Lubin P. 1993! ApJL! 412! L47-L50;

\refitem
26!
Gorski, K. 1992! ApJL! 398! L5-L8;

\refitem
27!
Lauer, T. \& Postman, M. 1992! BAAS! 24! 1264;

\refitem
28!
Wright, E. L. 1993! Annals of the New York Academy of Sciences! 688! 836; 

\refitem
29!
Crittenden, R., Bond, J. R., Davis, R. L., Efstathiou, G. \&
Steinhardt, P. J. 1993! PRL! 71! 324-327;

\refitem
30!
Wright, E. L., Smoot, G. F., Kogut, A., Hinshaw, G., Tenorio, L.,
Lineweaver, C., Bennett, C. L. \& Lubin, P. M. 1994! ApJ! 420! TBD; 

\refitem
31!
Martin, D. H. \& Puplett, E. 1970! Infrared Physics! 10! 105-109;

\refitem
32!
de Bernardis, P., Masi, S., Melchiorri, F., Melchiorri, B. \& Vittorio, N.
1992! ApJ! 396! L57-L60;

\refitem
33!
Watson, R. A., Gutierrez de la Cruz, C. M., Davies, R. D., Lasenby, A. N.,
Rebolo, R., Beckman, J. E. \& Hancock, S. 1992! Nature! 357! 660-665;

\bookref
34!
Dragovan, Mark, 1993, private communication.

\refitem
35!
Cheng, E. S., Cottingham, D. A., Fixsen, D. J., Inman, C. A., 
Kowitt, M. S.,
Meyer, S. S., Page, L. A., Puchalla, J. L. \& Silverberg, R. F. 
1993! ApJ! TBD! TBD;

\refitem
36!
Gundersen, J., Clapp, A., Devlin, M., Holmes, W., Fischer, M., 
Meinhold, P., Lange, A., Lubin, P. Richards, P. \& Smoot, G. 1993!
ApJ! 413! L1;

\refitem
37!
Meinhold, P., Clapp, A., Devlin, M., Fischer, M., Gundersen, J., 
Holmes, W.,
Lange, A., Lubin, P., Richards, P. \& Smoot, G. 1993! ApJL! 409! L1-L4;

\refitem
38!
Myers, S. T., Readhead, A. C. S., \& Lawrence, C. R. 1993! ApJ! 405! 8-29;

\refitem
39!
Readhead, A. C. S., Lawrence, C. R., Myers, S. T., Sargent, W. L. W.,
Hardebeck, H. E. \& Moffet, A. T. 1989! Ap. J.! 346! 566;

\refitem
40!
Fixsen \etal\ 1994! ApJ! 420! TBD;

\refitem
41!
Mather, J.~C. \etal\ 1990! ApJ! 354! L37-L41;

\refitem
42!
Wright, E. L. \etal\ 1991! ApJ! 381! 200-209;

\refitem
43!
Mather \etal\ 1994! ApJ! 420! TBD;

\refitem
44!
Fixsen \etal\ 1994! ApJ! 420! TBD;

\refitem
45!
Wright \etal\ 1994! ApJ! 420! TBD;

\refitem
46!
Walker, T. P., Steigman, G., Schramm, D. N., Olive, K. A. \& 
Kang, H-S. 1991!
ApJ! 376! 51-69;

\refitem
47!
Steigman, G. \& Tosi, M. 1992! ApJ! 401! 150-156;

\refitem
48!
Kompaneets, A. S.  1957! Sov. Phys. JETP! 4! 730;

\refitem
49!
Zeldovich, Ya. B. \& Sunyaev, R. A. 1969! Ap. \& Sp. Sci.! 4! 301-316;

\refitem
50!
Burigana, C., De Zotti, G. F., \&  Danese, L. 1991! ApJ! 379! 1-5;

\end{document}